\documentclass[]{aastex}
\usepackage{emulateapj5}

\newcommand{\etal}{{\it et al.}}
\newcommand{\msun}{{\rm M}_{\solar}}
\newcommand{\hmpc}{${h^{-1}}${\rm Mpc}}
\newcommand{\hkpc}{${h^{-1}}${\rm kpc}}
\newcommand{\hMsun}{{h^{-1}}{\rm M}_{\solar}}
\newcommand{\solar}{\ifmmode_{\mathord\odot}}
\newcommand{\LCDM}{$\Lambda$CDM}

\newcommand{\beq}{\begin{equation}}
\newcommand{\eeq}{\end{equation}}
\newcommand{\kms}{\, \rm{km}\,  \rm{s}^{-1}}
\newcommand{\kmsmpc}{\, \rm{km}\,  \rm{s}^{-1}\, \rm{Mpc}^{-1}}
\newenvironment{inlinefigure}{
\def\@captype{figure}
\noindent\begin{minipage}{0.999\linewidth}\begin{center}}
{\end{center}\end{minipage}\smallskip}

\lefthead{WECHSLER \etal}
\righthead{GALAXY FORMATION AT $Z \sim 3$ }
\begin{document}
\submitted{Received 2000 November 14; accepted 2000 December 13}
\journalinfo{The Astrophysical Journal, in press}

\title{Galaxy Formation at $z \sim 3$: Constraints from Spatial Clustering}

\author{Risa H. Wechsler\altaffilmark{1}, 
	Rachel S. Somerville\altaffilmark{2,4},
	James S. Bullock\altaffilmark{3,1}, 
	Tsafrir S. Kolatt\altaffilmark{4,1}, 
	Joel R. Primack\altaffilmark{1}, 
	George R. Blumenthal\altaffilmark{5}, \&
	Avishai Dekel\altaffilmark{4}
}

\altaffiltext{1}{Physics Department, University of California, Santa Cruz, 
	CA 95064; risa@physics.ucsc.edu, joel@ucolick.org}
\altaffiltext{2}{Institute of Astronomy, University of 
	Cambridge, Madingley Road, Cambridge, CB3 0HA, United Kingdom;
	rachel@ast.cam.ac.uk}
\altaffiltext{3}{Department of Astronomy, The Ohio State 
	University, 140 W. 18th Ave, Columbus, OH 43210;
	james@astronomy.ohio-state.edu}
\altaffiltext{4}{Racah Institute of Physics, The Hebrew 
	University, Jerusalem 91904 Israel;
	tsafrir@astro.huji.ac.il, dekel@astro.huji.ac.il}
\altaffiltext{5}{Astronomy \& Astrophysics Department, University of 
	California, Santa Cruz, CA 95064; george@ucolick.org}

\begin{abstract}
We use N-body simulations combined with semi-analytic models to
compute the clustering properties of modeled galaxies at $z\sim 3$,
and confront these predictions with the clustering properties of the
observed population of Lyman-break galaxies (LBGs). Several scenarios
for the nature of LBGs are explored, which may be broadly categorized
into models in which high-redshift star formation is driven by
collisional starbursts and those in which quiescent star formation
dominates.  For each model, we make predictions for the LBG
overdensity distribution, the variance of counts-in-cells, the
correlation length, and close pair statistics.  Models which assume a
one-to-one relationship between massive dark-matter halos and galaxies
are disfavored by close pair statistics, as are models in which
colliding halos are associated with galaxies in a simplified way.
However, when modeling of gas consumption and star formation is
included using a semi-analytic treatment, the quiescent
and collisional starburst models predict similar clustering properties
and none of these models can be ruled out based on the
available clustering data.  None of the ``realistic'' models predict a
strong dependence of clustering amplitude on the luminosity threshold
of the sample, in apparent conflict with some observational results.  
\end{abstract}


\section{INTRODUCTION}
In recent years there has been impressive growth in observations of
high-redshift galaxies. The ``Lyman-break'' technique
\citep{steidel:92,madau:96,steidel:96a} makes it possible to select
high-redshift candidates based on their photometric colors. Extensive
spectroscopic follow-up has confirmed that this technique very
reliably selects high redshift ($z \ga 2$) galaxies
\citep{steidel:96a,steidel:96b,low:97}.  The largest sample covers the
redshift range $2 \la z \la 3.5$, where over 1200 photometric
candidates and about 900 spectra have now been obtained, mainly by
Steidel and collaborators.  Similar techniques can be used to identify
galaxies at even higher redshifts, although spectroscopic confirmation
is more difficult.  About fifty confirmed objects exist at $4.5
\la z \la 5.5$ \citep{steidel:99} and a handful at $z \ga 5.0$
(e.g., \citealt{weymann:98,spinrad:98}).  Our main focus in this paper
will be the $z\sim3$ LBG sample accumulated by the Steidel group,
which is fairly complete to ${\mathcal R} = 25.5$, allowing robust
estimation of the clustering properties at this redshift and magnitude
limit.

The correlation length of the $z\sim3$ sample is similar to that of
nearby bright galaxies ($r_0 \sim 3$--6 \hmpc, comoving;
\citealt{adel:98,giav:98,giav:00}; hereafter A98, G98 and G00). 
Within the Cold Dark Matter (CDM) hierarchical structure formation
paradigm, these galaxies must therefore be much more clustered than
the underlying dark-matter density field (i.e., strongly
``biased''). Moreover, because the clustering of matter increases
monotonically with time, the bias of the Lyman-break galaxies
must be significantly higher than that of typical galaxies at
$z=0$. Although the actual level of bias and the details of its
redshift dependence depend on the cosmological model and the sample
selection, qualitatively this result is quite general and was pointed
out by the first observational papers on LBG clustering (A98,G98) as
well as numerous subsequent works. This is clearly a key property of
high-redshift galaxies and must be explained by any successful theory
of galaxy formation.

In the CDM framework, given a power spectrum and a cosmology, the
clustering properties of dark-matter halos can be readily estimated,
either by analytic methods or using N-body simulations. Numerous groups have
shown, using a variety of methods, that the observed clustering and
high bias of high-redshift galaxies can plausibly be reproduced in a
broad range of CDM cosmologies
\citep{mofuku:96,adel:98,wech:98,jing:98,bagla:98a,baugh:98,gov:98,coles:98,
moscar:98,khw:99,arnouts:99,kauf:99b,blanton:00}.  This implies that
the clustering properties of LBGs are not likely to provide very
discriminatory constraints on cosmology, especially as long as secure
knowledge about their masses is lacking. However, there is still hope
that LBG clustering may provide important constraints on galaxy
formation.

A remaining central uncertainty is the association of dark-matter
halos with observable galaxies.  Many previous investigations
\citep{mofuku:96,adel:98,wech:98,jing:98,bagla:98a,coles:98,moscar:98,arnouts:99}
have made the simple assumption that every dark-matter halo above a
given mass threshold hosts one observable LBG, and that the galaxy
luminosity is closely connected with the mass of the host halo.
Within the observational uncertainties at the time of publication of
these earlier works, the observed number density and correlation
length of the $z \sim 3$ sample could be reproduced within this sort
of scenario, provided that LBGs were associated with massive halos
($\ga {\rm few} 10^{11} \hMsun$ for low-$\Omega$ cosmologies). We
shall refer to this class of models as ``Massive Halo'' models for the
remainder of this paper.

\citet[hereafter K99]{kolatt:99} investigated a very different
model for LBGs, one in which \emph{all} of the observed high-redshift
galaxies are visible because they are temporarily brightened by
starbursts triggered by collisions. Colliding halos were identified in
a high-resolution N-body simulation, and a simple approach was used to
associate these collisions with visible LBGs.
Collisions between subhalos can lead to multiple LBGs 
within the same virialized halo. Although many of the
objects in this scenario are far less massive than in the Massive Halo
scenario described above, K99 showed that the correlation length of
the colliding halos was comparable to that of the observed LBGs, and
that the colliding halos were biased with respect to the dark matter.
This demonstrated that the mass threshold of the host halos does not
uniquely determine the clustering properties of a population of
objects. We shall investigate a model similar to the K99 model, which
we refer to as the ``Colliding Halo'' (CH) model.

Though both the Massive Halo and Colliding Halo models were able to
simultaneously fit the number density and clustering properties of
LBGs, they both rely on an ad hoc connection between dark-matter halos
and observable galaxies, and are almost certainly too simple to be
correct in detail. More detailed modeling of LBGs, relying on either
semi-analytic modeling or hydrodynamic simulations to treat the
physics of gas cooling and star formation, has led to a variety of
different views regarding the masses and basic nature of the LBG
population.  Using a semi-analytic model similar to that presented by
\citet{cafnz:94}, \citet{baugh:98} showed that under their
assumptions, LBGs are hosted by massive halos ($\ga 10^{12} \hMsun$),
and are forming stars mainly quiescently at a moderate rate.  The
correlation length of LBGs in their model was similar to that obtained
in the simpler Massive Halo models and consistent with the
observational estimates available at that time (see also
\citealt{gov:98}). We refer to this picture, in which LBGs are
massive, quiescently star-forming objects, as the ``massive
quiescent'' scenario.

Also using semi-analytic models, \citet*[hereafter SPF]{spf:00} showed
that the numbers and properties of high-redshift galaxies in such
models are very sensitive to the star formation recipe adopted. They
investigated three models, corresponding to three different recipes
for star formation, all of which produced good agreement with local
observations. In the ``Constant Efficiency Quiescent'' (CEQ) model,
all star formation occurs in a quiescent mode and the star formation
efficiency (i.e. the star formation rate per unit mass of cold gas) is
constant with redshift.  In the ``Accelerated Quiescent'' (AQ) model,
all star formation is quiescent but its efficiency scales inversely
with the disk dynamical time, thus increasing rapidly at high
redshift. In the third, the ``Collisional Starburst'' (CSB) model, in
addition to quiescent star formation, galaxy-galaxy mergers (both
major and minor) are assumed to trigger starbursts --- brief episodes
in which the rate of star formation is dramatically higher than in the
usual quiescent mode.
The Collisional Starburst model was favored by SPF as they found that
it produced the best overall agreement with the high-redshift data they
investigated.

Based on hydrodynamic simulations, \citet{khw:99} and
\citet{weinberg:00} supported a view intermediate to the massive
quiescent scenario of \citet{baugh:98} and the Collisional Starburst
scenario favored by SPF, although closer to the first.
\citet{weinberg:00} found that their simulated LBGs resided within
halos with a wide range of masses, but they still reproduced the
strong clustering observed.  Most of the LBGs in their simulations do
not appear to be undergoing starbursts, but the simulations do not
have sufficient mass or spatial resolution to properly treat most of
the collisions that SPF found to be important.

It is clear that regardless of whether semi-analytic or numerical
techniques are used, the results of theoretical predictions about the
nature of LBGs depend sensitively on the highly uncertain physics of
star formation and feedback. Each of the proposed scenarios has
potential problems. The simple Massive Halo models and the more
detailed massive quiescent-type models seem to reproduce the observed
clustering strength of LBGs, but the ``realistic'' versions of these
models --- e.g., the Constant Efficiency Quiescent model of SPF ---
have difficulty producing enough objects when dust extinction is
included, and predict that the number density of bright galaxies
should decline rapidly at higher redshift, in apparent conflict with
observations (SPF).  An alternative recipe for quiescent star
formation --- the Accelerated Quiescent recipe of SPF --- gives
acceptable agreement with the number density of LBGs at $z \ga 3$.
However, this model has difficulty in producing enough very bright
objects, and also consumes so much gas that it violates constraints
from observations of Damped Lyman-$\alpha$ systems (SPF)
\footnote[1]{The star formation recipe used in the hydrodynamic
simulations is similar to the AQ model of SPF, since the gas
consumption timescale scales with the local dynamical time.  The mass
resolution is not good enough to tell conclusively whether the large
amount of high-redshift gas consumption results in the same problem
with matching the DLAS abundance, but \citet{gardner:99} argue based
on an analytic extention of the mass resolution that this is not a
serious problem.}.  In addition, because LBGs are found in smaller
mass halos, it was not clear whether they would be clustered enough to
match the data.  The Colliding Halo model of K99 was shown to
reproduce the clustering of LBGs on scales of several Mpc, but it may
be {\em too} clustered on smaller scales, and thus overpredict the
number of close pairs \citep{mmw:99}. The clustering properties of the
more detailed Collisional Starburst model of SPF have not been checked
until the present work, but could suffer the same problem.  Also,
there is a suggestion that the clustering strength of observed LBGs
depends on the luminosity threshold of the sample
(\citealt{steidel:98rs}, hereafter S98; G00), with brighter galaxies
being more strongly clustered. Because of the expected large
dispersion in the relationship between mass and luminosity, starburst
models might have difficulty producing a strong trend of this sort.

The goal of this paper is to test a set of models covering the full
range of previously proposed scenarios for the nature of LBGs, from
the very simple Massive Halo and Colliding Halo models to the more
``realistic'' models mentioned above, and to determine which of them, if
any, can be ruled out by comparing their predicted clustering
properties with the available data at $z\sim 3$. The number of
observable objects per halo (the occupation function) is calculated
using both simple analytic prescriptions and results from the
semi-analytic models of SPF.  Large-volume dissipationless N-body
simulations are used to calculate the expected clustering properties
of halos at $z=3$, and the calculated occupation functions are used to
convert this into predictions for the clustering properties of
observable galaxies (this is similar to the approach used by
\citealt{kns:97} and
\citealt{benson:00}).  
We mimic the observational selection effects as closely as possible,
apply them to model galaxies, and then compare these predictions to
the data in the ``observational plane''.  Throughout the paper, we
focus on one cosmology, the currently popular
\LCDM\ model, with matter density $\Omega_m = 0.3$, vacuum energy
density $\Omega_{\Lambda} = 0.7$, and a Hubble parameter $h=0.7$,
where $H_0 = 100h \kmsmpc$.

The outline of the paper is as follows.  We begin (\S\ref{sec:mw-lc})
with an analytic investigation of clustering for two extreme models
for LBGs: Massive Halos and Colliding Halos. In \S\ref{sec:data}, we
discuss the data that we will use for comparison. In \S \ref{sec:sims}
we discuss the N-body simulations we use to derive halo clustering
properties, and the models that are used to populate these halos with
galaxies.  In \S \ref{sec:results}, we present the statistics used and
the results of our comparison.  In \S \ref{sec:compare}, we compare
the Colliding Halo model and the more detailed semi-analytic
Collisional Starburst model and determine which elements of the models
are responsible for differences in their behavior. We discuss our
results and conclude in \S \ref{sec:conclu}.

\section{HALO OCCUPATION AND CLUSTERING}
\label{sec:mw-lc}

In this section we explore the clustering properties of two toy
models representing opposite extremes of the spectrum of proposed
scenarios for the nature of LBGs: the Massive Halo model, in which
LBGs are associated in a one-to-one fashion with the most massive
halos, and the Colliding Halo model, in which LBGs are associated with
collisions between halos and/or subhalos. In several previous works
(for example A98), the observed clustering of LBGs has been used to
obtain estimates of the characteristic masses of their host halos. As
shown below, an additional factor in the expected clustering of any
population of objects is the average number of objects residing within
dark halos of a given mass (the halo occupation function). The unknown
occupation function introduces a degeneracy which results in a
significant uncertainty in the minimum host halo mass corresponding to a given
clustering strength.

For each of our toy models, the halo occupation function $N_g(M)$ is
approximated as a power-law of the mass for halos larger than some
minimum mass $M>M_{\rm min}$: $N_g(M)\propto M^S$. Here $M_{\rm min}$
corresponds to the minimum mass halo capable of hosting an observable
galaxy\footnote[2]{Exactly what is meant by an ``observable'' galaxy
obviously depends on the particular techniques used and the redshift,
bandpass, and sensitivity limit of a given sample. In this paper, we
focus on the ground-based spectroscopic sample of $U_n$ drop-outs
($\bar{z} \sim 3$) of Steidel et al., which has a magnitude limit of
approximately ${\cal R}_{AB}$ = 25.5. We have these objects in mind
when referring to ``observable'' galaxies.}.  The Massive Halo model
has the simple form $S=0$ --- each halo above some minimum mass is
assumed to host exactly one galaxy.

For the Colliding Halo model, a simple approximation for the slope of
the occupation function can be obtained using the following argument.
Assume that the number of collisions that occur within a halo of mass
$M$ over a time interval $\Delta t$ is proportional to the amount of
mass that halo has accreted during this time interval divided by the
average mass of the accreted objects:
\begin{equation}
 N_{\rm coll} \propto \Delta M/<M_{ac}>. 
\end{equation}

For $\Delta t \ll t$, where $t$ is the age of the universe at the time
the halo is observed, we use the single-trajectory formula of
\citet{lacey:93} in order to estimate the median amount of mass
$\Delta M$ accreted by a halo of mass $M$ in a time period $\Delta t$:
\begin{equation}
0.5 = {\rm erfc}\left[\frac{\delta_c(t) - \delta_c(t + \Delta t)}{\sqrt{
     2[\sigma^2(M) - \sigma^2(M + \Delta M)]}} \right],
\end{equation} 
where $\delta_c(t)=\delta_{c, 0} /D(t)$ is the linearly extrapolated critical density, in which $\delta_{c, 0}\simeq 1.68$
and $D(t)$ is the linear growth factor. The quantity $\sigma(M)$ is 
the linear rms fluctuation inside a spherical window of mass $M$, also
equal to the square root of the mass power spectrum.  If we approximate 
it as a power law 
$\sigma(M)
\propto M^{- \alpha}$, and assume $\Delta M \ll M$, we find
\begin{equation} \Delta M \propto M^{1 + 2\alpha}.
\end{equation}

To obtain a rough estimate of the average accreted mass, we
approximate the mass spectrum of accreted halos with the power-law
form $dN/dM_{ac} \propto M_{ac}^{\alpha-2}$ for $M_{ac} \ll M$
\citep{lacey:93, ps:74}.  For $M_{\rm min} \ll \Delta M$ this implies
\begin{equation}
<M_{ac}> \simeq \int_{0}^{\Delta M} \frac{dN}{dM_{ac}} M_{ac} dM_{ac} 
\propto \Delta M^{\alpha}.
\end{equation}
Equations 1, 3 and 4 then imply
\begin{equation}
N_{\rm coll} \propto M^{1 + \alpha - 2 \alpha^2}.
\end{equation}

For our \LCDM\ cosmology at a mass scale of $M \simeq 10^{12} \hMsun$,
$\alpha \simeq 0.15$, leading to a value of $S \simeq 1.1$.  In \S
\ref{sec:sims:ch}, we show that this is in reasonable agreement with the
results from high-resolution N-body simulations, and with results from
semi-analytic Monte Carlo merger-trees. Of course, the approximation
should break down for small halo masses $M \simeq M_{\rm min}$.

With simple expressions for the halo occupation function in hand, it
is now straightforward to calculate the clustering properties of each
model. In the first case, since LBGs are found only in the most
massive halos, their clustering will be biased with respect to the
underlying dark-matter distribution (see e.g., W98, A98). In the second
case, although collisions can be found in smaller-mass halos, they
will preferentially be located within large halos, and the
distribution of collisions should also be strongly biased.

The bias $b$ is defined here as a relation between the correlation
functions of halos and dark matter: $b_{\rm h}^2 = \xi_{\rm
h}/\xi_{\rm DM}$.  For halos of mass $M$ at redshift $z$ and on scales
large compared to the size of collapsed halos, an approximate
expression for the bias is given by \cite{mw:96}:
\begin{equation}
b_h(M,z) =1+\frac{\nu(M, z)^2-1}{\delta_c(0)},
\label{eqt:mw}
\end{equation}
where $\nu(M, z) = \delta_c(z)/\sigma(M)$.

To do the numerical calculations in this section, we 
use the expression given by
\citet{jing:99}:
\begin{equation}
b_h(M,z) = \left( \frac{0.5}{\nu(M, z)^4} +1 \right) 
\left(1+\frac{\nu(M, z)^2-1}{\delta_c(0)} \right),
\label{eqt:jing}
\end{equation}
a modification of \citet{mw:96} which produces better agreement with
N-body simulations.

The appropriate bias factor for a population of galaxies within halos
more massive than $M_{\rm min}$ is the average of $b_h(M,z)$ weighted
by the abundance of halos as a function of mass, $dN_h/dM$ (e.g., as
estimated by the approximation of \citealt{ps:74}; here we use the
expression given by \citealt{st:99} which again is a better fit to
simulations), and the average number of LBGs per halo $N_g(M)$:
\begin{eqnarray}
\lefteqn{b_g(z, M>M_{\rm min}) =} \nonumber \\
&& \frac{1}{N_g(z)} \int_{M_{\rm min}}^{\infty}
\frac{dN_h}{dM}(M,z) b_h(M,z) N_g(M) dM,
\label{eqt:bg}
\end{eqnarray}
where $N_g(z) = \int_{M_{\rm min}}^{\infty} \frac{dN}{dM}(M,z) N_g(M)
dM$.  For the Massive Halo model, $N_g(M)= M^{0}=1$ and the standard
expression for the average bias of halos (averaged over a particular
mass range) obtains.  As above, we use $N_g(M)\propto
M^{1.1}$ to represent the toy Colliding Halo model.  

\begin{inlinefigure}
\begin{center}
\resizebox{\textwidth}{!}{\includegraphics{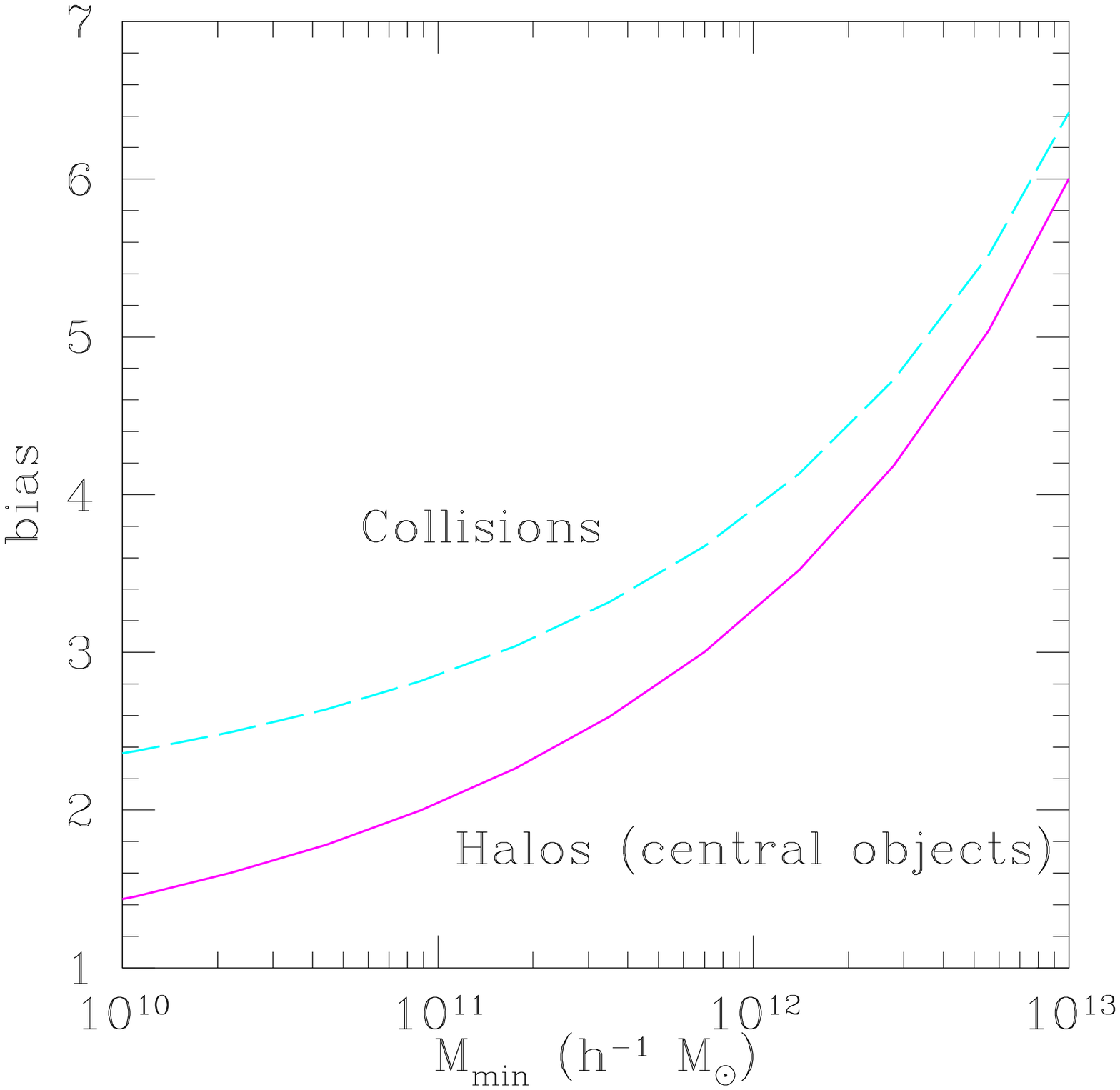}}
\end{center}
\figcaption{Bias parameter at z$\sim 3$ (using Equation 7) for halos 
   [$N_g(M) = 1$] and collisions [$N_g(M) \propto M^{1.1}$]
as a function of the minimum
host halo mass $M_{\rm min}$.
\label{fig:bias}}
\end{inlinefigure}

In Figure \ref{fig:bias}, the galaxy bias is plotted as a function of
the minimum host halo mass at $z=3$ for both toy models, assuming our
usual \LCDM\ cosmology.  Because high-mass halos are weighted more strongly
in the Colliding Halo model, galaxies are more biased for fixed $M_{\rm
min}$:  the observed bias for LBGs in this cosmology ($b
\sim 2-3$) corresponds to values of $M_{\rm min} \sim 10^{11} \hMsun$
for the Colliding Halo model, versus $M_{\rm min} \sim 10^{12} \hMsun$
for the Massive Halo model.  This was also shown using N-body
simulations by \citet{kolatt:99}, and is discussed further in \S
\ref{sec:results:cf}.

Note that the above discussion pertains to clustering on scales larger
than the sizes of virialized halos. Clustering on smaller scales is
extremely sensitive to the slope of the occupation function $N_g(M)$,
and is discussed further in \S \ref{sec:results:pairs}.  As a final
aside, we note that the steep slope of the occupation function for
collisions is relevant when estimating the clustering properties of
any population believed to be associated with mergers, such as quasars
or AGN \citep[e.g.,][]{haiman:00,martini:00}.  In the next section we discuss
detailed predictions for a number of models for populating halos
within an N-body simulation.

\section{DATA}
\label{sec:data}
Steidel, Adelberger, and their collaborators have compiled a large
sample of bright galaxies at high redshifts \citep[S98;
A98;][]{giav:98, adel:01}. The sample of photometric candidates (based
on $U_n$, ${\cal G}$, \& ${\cal R}$ photometry) now consists of
roughly 1300 objects in 15 fields. Spectroscopic redshifts have been
obtained for about half this number, and have confirmed that this
technique very reliably selects objects in the redshift range $2.2\leq
z \leq 3.8$, with a median redshift of $\bar{z}\sim 3$. The
photometric sample is estimated to be fairly complete at ${\cal R}
\leq 25.5$ (though it remains uncertain whether a significant fraction
of the true high-redshift population is missed because the colors lie
outside the photometric selection area, for example due to extreme
dust reddening; see the discussion in \citealt{as:00} and references
therein).  

In the present analysis, we make use of data from a sample of 500
galaxies with spectroscopic redshifts.  Most of these data are
published in A98, who found 376 galaxies in this redshift range, in
six $9\arcmin \times 9 \arcmin$ fields. Also included here is an
analysis of an additional two fields of the same size provided to us
by K.  Adelberger.  In order to perform a fair comparison with
theoretical predictions, some assumptions must be made about how the
observed galaxies are selected.  We assume that the true comoving
number density of galaxies is constant over the redshift range $2.5
\leq z \leq 3.5$, and that the selection function over this redshift
range is given by the fit to a histogram of all A98 data.  At the peak
of the selection function, ($z \sim 3$), we assume that $\sim 70\%$ of
all galaxies with ${\cal R} \leq 25.5$ would be identified as
photometric candidates (this completeness percentage is still somewhat
uncertain, as mentioned above; we choose it so that we match the most
recent estimate of the incompleteness-corrected number density given
by \citealt{adel:99}, see below).  Spectroscopic redshifts are
successfully obtained for $40\%$ of the photometric candidates in this
sample; for simplicity (and since the dependence is not yet fully
understood) we ignore the probable tendency of the spectroscopic
sample to preferentially include brighter galaxies.  In addition, the
selection function falls off on either side of $z \sim 3$.  This
implies that the true number density of LBGs in this redshift range
with ${\cal R} \leq 25.5$ is about 0.004 h$^3$ Mpc$^{-3}$, which is
roughly seven times the number density of LBGs with measured redshifts
(note that no attempt is made to correct the observations for dust
extinction; instead we will apply dust corrections to the theoretical
models).

The statistics that we will use to compare with our models are the
distribution of overdensities and the variance of counts in cells of
roughly 12 \hmpc\ on a side, the correlation function, and the
fraction of galaxies in pairs within $1 \arcmin$. The first two
quantities are calculated for the spectroscopic sample by A98, and the
correlation length is calculated for the photometric sample by G98 and
G00. The close pair data were provided to us by K. Adelberger. We also
investigate the dependence of the correlation length on the magnitude
limit of the sample, which has been discussed in S98 and G00.

\section{MODELING}
\label{sec:sims}
\subsection{Halo Clustering: N-Body Simulations}
\label{sec:sims:nbody}
Cosmological N-body simulations are used to obtain the spatial locations
and masses of virialized dark-matter halos at $z=3$. The simulations
were produced by the GIF collaboration\footnote[3]{Performed at the
Max-Planck-Institut f\"{u}r Astrophysik, Garching, and the Edinburgh
Parallel Computing Centre using codes from the Virgo Supercomputing
Consortium
(\url{http://star-www.dur.ac.uk/$^{\sim}$frazerp/virgo/virgo.html}); see
\citet{jenk:98} for a discussion of these codes and of related
simulations. The halo catalogs used here are now publicly available at
http://www.mpa-garching.mpg.de/NumCos}. Only one cosmology is
considered here, a flat \LCDM\ model with $\Omega_m = 0.3$, $h = 0.7$,
$\sigma_8$ = 0.9, and a shape parameter $\Gamma=0.21$.  The box is 141
\hmpc\ on a side, and the simulation includes $256^3$ particles of mass $1.4 \times 10^{10}
{h^{-1}}{\rm M}_{\mathord\odot}.$  
Virialized halos were identified using a standard
friends-of-friends algorithm and only halos with at least 10 particles
were used for our analysis (see \citealt{kauf:99a} for a more detailed
description of the halo catalogs).

\subsection{Populating Halos with Galaxies}
\label{sec:sims:pop}
Five different models are considered for populating these halos with
observable galaxies. The first two models associate galaxies with
either massive halos or halo collisions using very simple, ad hoc
prescriptions. For the second set of models, detailed semi-analytic
modeling is used in an attempt to calculate the number of observable
galaxies per halo from a ``forward evolution'' approach. The three
models in this set correspond to the different recipes for quiescent
and bursting star formation considered by SPF.  The models are
summarized in Table~\ref{tab:models}, and described in more detail
below.  Each model is normalized so that the underlying population of
galaxies brighter than  ${\cal R}=25.5$ has a number density of 0.004
h$^3$ Mpc$^{-3}$.  The parameters used to obtain this normalization
are different for each type of model, and are discussed further below.

Once the halos have been populated with galaxies, we try to mimic the
observational selection process by creating an ``observed'' sample of
galaxies according to the assumptions outlined in \S\ref{sec:data}.
The simulation box is broken into pixels the size of the data fields,
and the galaxies in each pixel are observed according to a selection
probability, randomly chosen from one of the data pixels.  The
resolution of the ground-based images is about $1-2 \arcsec$
(K. Adelberger 1999, private communication), so model galaxies within
$\sim 1.5 \arcsec $ of each other are treated one galaxy.
\\
\subsubsection{Massive Halos}
\label{sec:sims:mh}
In the simplest possible model for LBGs (the Massive Halo model),
each halo more massive than a given threshold is assumed to host
exactly one LBG. This minimum mass comprises the one adjustable
parameter of the model, and is chosen to obtain the observed number
density of LBGs.  Similar models have been considered previously by
many authors
\citep[e.g., W98;][]{adel:98,jing:98,coles:98,moscar:98,bagla:98a,arnouts:99}.

\subsubsection{Colliding Halos}
\label{sec:sims:ch}
The Colliding Halo model is a simple representation of the idea
that galaxies may be made visible by short episodes of star formation
triggered by collisions. This model is based on the analysis of a
high-resolution N-body simulation which uses the ``Adaptive Refinement
Tree'' (ART) algorithm \citep{kkk:97} to obtain very high force
resolution ($\sim 1 h^{-1}$kpc) in a $30 h^{-1}$Mpc box.  The
simulation we use is for the same \LCDM\ cosmology mentioned
previously except that here $\sigma_8=1.0$ instead of $0.9$. Halo
catalogs were created using a variant of a spherical overdensity
method \citep{bullock:00} which was explicitly designed to allow the
identification of subhalos located within the virial radius of larger
halos. Halo and subhalo collisions were then identified using the
approach described in \citet{kolatt:99}.  The mass per dark-matter
particle is $1.25\times10^{8}\hMsun$ and the halo catalogs are
complete for halos more massive than $\sim 2\times 
10^{10} \hMsun$ \citep{sigad:00}. 

The small volume of the ART box does not allow us to robustly
calculate some of the clustering statistics directly.  The occupation
function of collisions is therefore determined by assigning each
identified collision to the host halo that it resides in at the end of
the timestep. Figure \ref{fig:pl} shows this result for a timestep
covering $2.9<z<3.9$, as well as for the same time interval divided
into two sub-steps. The average number of collisions per halo as a
function of halo mass is very well represented by a power law $N_{\rm
coll} \propto M_{\rm host}^S,$ with a value of $S
\simeq 1.13$.  This is very close to the power-law slope predicted by
the analytic argument in \S \ref{sec:mw-lc}.  Other timesteps
exhibit a similar power law, as does a larger (60 \hmpc), lower
resolution box.

This power law is now used to populate the dark-matter halos in the
larger, lower resolution GIF simulations with galaxies.  We assume
that the minimum mass for a halo to host a collision producing a
visible LBG is $M_{\rm min} = 10^{11} h^{-1}M_{\odot}$.  The
normalization of the power-law is set by requiring the total number of
objects to be the same as the observed density of LBGs.  Note
that there are sufficient collisions to account for the required normalization
in the simulation.  For each
halo, the actual number of objects is chosen from a Poisson
distribution with the mean given by the power-law function: $N =
CM^{1.1}, M>M_{\rm min}$.  The sensitivity of our results to the
modeling of scatter is discussed in the following section.

\begin{inlinefigure}
\resizebox{\textwidth}{!}{\includegraphics{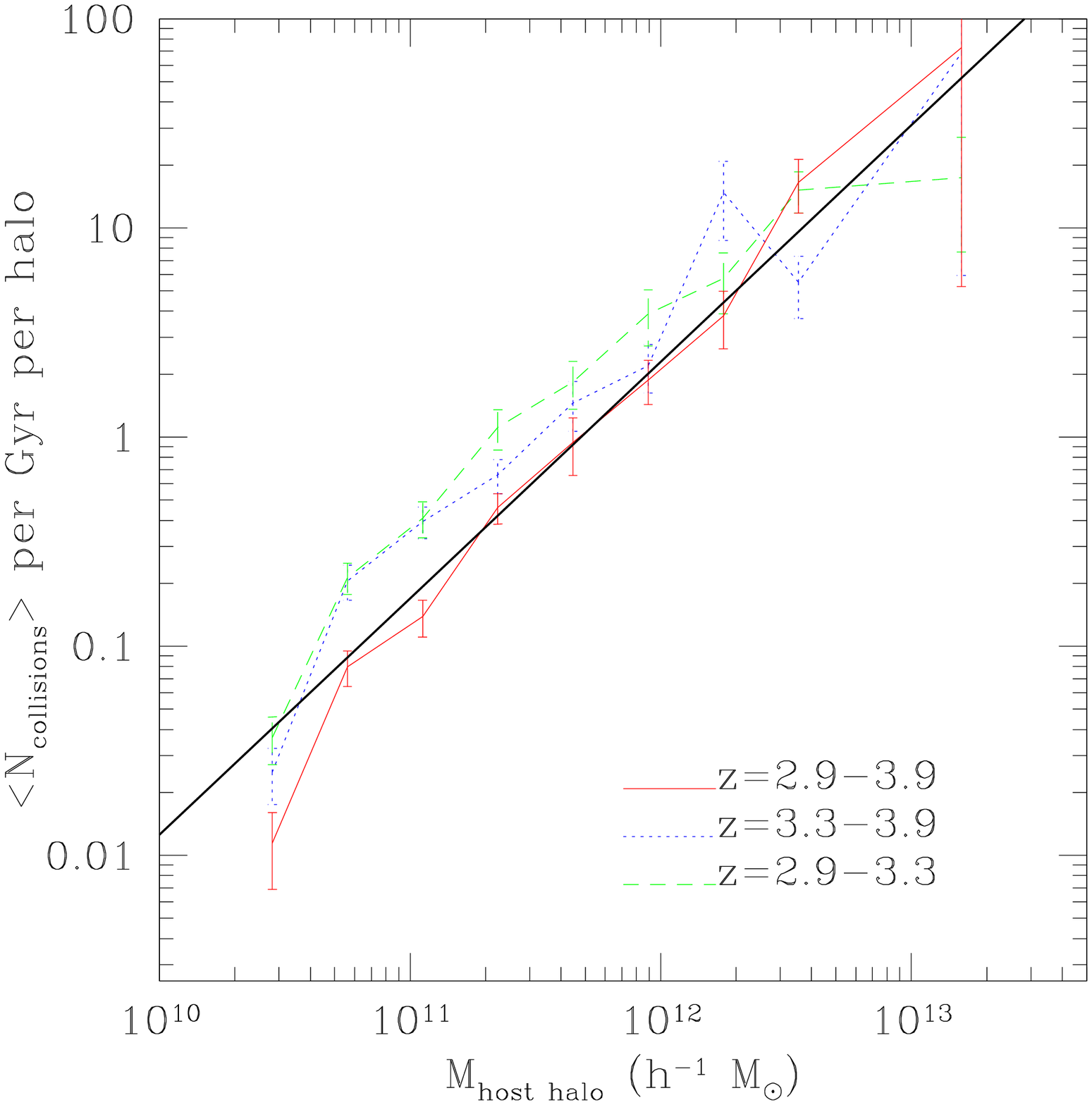}}
\figcaption{Average number of collisions per halo from the 30 \hmpc\
ART simulation, for three high-redshift timestep intervals.  
The error bars shown are
1-$\sigma$ scatter in this value.  The best fit slopes for these lines
are $1.28 \pm 0.07$, $1.03 \pm 0.13$, $1.07 \pm 0.09$, for the
intervals $z=2.9-3.9$, $z=3.3-3.9$, $z=2.9-3.3$, respectively.  The
heavy line shown is the weighted average for the three timesteps, which
yields a slope of 1.13 --- similar to the power-law slope value derived
from an analytic argument in \S \ref{sec:mw-lc}. 
\label{fig:pl}}
\end{inlinefigure}

\subsubsection{Semi-Analytic Models}
\label{sec:sims:sams}
The Massive Halo model and the Colliding Halo model are not much more
than toy models, normalized by adjusting ad hoc parameters.
Predicting the number of galaxies within a halo and their luminosities
from first principles is a rather daunting proposition. Semi-analytic
models attempt to capture the complex interplay of the physics of
gravitational collapse and merging, gas dynamics, and star formation
and feedback, by using simple recipes to model each of these physical
processes.  The semi-analytic models used here were developed by
\citet{rsthesis} and are described in SP and SPF.  Here we give a
brief description of the models, emphasizing the aspects most relevant
to the present analysis.  The reader is referred to SP, SPF, and
references therein for further details.

The formation and merging of dark-matter halos as a function of time
is represented by a ``merger tree'', which is constructed using the
method of \citet{sk:99}. Halos with velocity dispersions less than
$\sim 40 \kms$ are assumed to be photo-ionized so that the gas within
them cannot cool or form stars. This sets the effective mass
resolution of our merger trees. When halos merge, the central galaxy
in the largest progenitor halo becomes the new central galaxy and all
other galaxies become satellite galaxies orbiting within the halo.
Satellite galaxies fall towards the center of the halo due to
dynamical friction and eventually merge with the central galaxy.
Satellite galaxies may also merge with each other according to the
modified mean free path model of \citet[see SP \& SPF for
details]{mh:97}.

When a halo collapses, the gas within it is assumed to be shock heated
to the virial temperature of the halo.  This gas is transformed to
``cold'' gas when the time elapsed since the halo collapsed is equal to
the time needed for it to radiate away all of its energy. This
``cooling time'' depends on the density, temperature, and metallicity of
the hot gas.

Quiescent star formation occurs in all disk galaxies that possess cold
gas, according to the expression 
\begin{equation}
\dot{m}_*={m_{\rm cold}\over{\tau_*}} , 
\end{equation}
where $m_{\rm cold}$ is the mass in cold gas and $\tau_*$ is the ``star
formation timescale'', which is a parameterization of our ignorance
about star formation. SP and SPF considered two cases for quiescent
star formation, ``constant efficiency'', in which $\tau_*$ is
constant, and ``accelerated'', in which $\tau_* \propto t_{\rm dyn}$,
where $t_{\rm dyn}$ is the dynamical time of the disk (this is similar
to the recipe used by e.g., \citealt{kauf:99a}). The accelerated
recipe is so-named because disk dynamical times are smaller at earlier
times, leading to a dramatic increase in the star formation efficiency
with redshift.  Other authors have considered recipes in which
$\tau_*$ depends explicitly on circular velocity
\citep{cafnz:94,baugh:99}.

In addition, when galaxies merge, a ``burst'' mode of star formation
may be triggered. The recipe for star formation in bursts adopted by
SPF was an attempt to parameterize the results of hydrodynamical
simulations of pairs of colliding galaxies
\citep{mihos:94,mihos:95,mihos:96}. In a series of papers, 
Mihos \& Hernquist investigated both major (mass ratio 1:1) and minor
(mass ratio 1:10) mergers. They found that major mergers typically
triggered a burst which consumed 65-80 percent of the available cold
gas over several hundred Myr, whereas a minor merger between a
satellite and a pure disk galaxy consumed 30-50 percent of the gas
over a similar timescale. However, if the larger galaxy possessed a
bulge of one-third the disk mass, the burst was suppressed in the
minor merger case, and only about 5 percent of the gas was
consumed. To attempt to represent this behavior, SPF modeled the
burst efficiency (the fraction of cold gas consumed during the
burst) as a power-law function of the mass ratio of the merger:
\begin{equation}
e_{\rm burst} = \left (\frac{m_{\rm small}}{m_{\rm big}}\right)^{\alpha} \, ,
\label{eqn:eburst}
\end{equation}
where the adopted value of $\alpha=0.18$ for the no-bulge case (in
which the bulge mass is less than one-third of the disk mass) and
$\alpha=1.18$ for the bulge case were chosen to match the two cases
simulated by Mihos \& Hernquist.  We comment later on uncertainties in
these parameters, which were all based on simulations of collisions of
galaxies which initially resemble low-redshift galaxies.  In SPF, the
burst timescale was assumed to be equal to the disk dynamical time,
which is probably a lower limit on the burst
timescale\footnote[4]{\citet{kennicutt} finds that the gas consumption
times for starburst galaxies are generally smaller than, and never
exceed, their dynamical times.}.

Chemical evolution is modeled assuming that each generation of stars
produces a fixed yield of metals. These metals are initially deposited
in the cold gas, and may be subsequently mixed with the hot halo gas,
or ejected from the halo, by supernovae feedback.
The luminosity of each galaxy at the desired redshift and in the
desired bands is then calculated using stellar population synthesis
models. Here we have used the most recent version of the models of
Bruzual \& Charlot (GISSEL00), and assumed a solar metallicity SED and
a Salpeter IMF. We have checked that the results of the GISSEL00
models are consistent with the 1998 versions used in SPF, and that the
results presented here are not sensitive to the assumed metallicity of
the stellar population.

The semi-analytic models contain a number of free parameters, with the most
important being the three that govern the efficiency of quiescent star
formation, the efficiency of supernovae feedback, and the yield of
metals per solar mass of stars produced. These parameters are set by
requiring an average ``reference galaxy'' (with $V_c = 220$ km/s) at
redshift zero to have the correct luminosity, gas content, and
metallicity, as specified by observations of nearby galaxies (see SP
for details).

We shall investigate the same three models considered by SPF, which
differ only in the treatment of star formation:
\begin{enumerate}
\item {\it Constant Efficiency Quiescent (CEQ)} : quiescent star
formation only (no bursts), and $\tau_* \equiv m_{\rm
cold}/\dot{m}_{*} =$ constant.

\item {\it Accelerated Quiescent (AQ)} : quiescent star formation only
(no bursts), and $\tau_* \equiv m_{\rm cold}/\dot{m}_{*} \propto
t_{\rm dyn}$. For a given halo mass, $t_{\rm dyn}$ is smaller at high
redshift because collapsed objects are denser, therefore a given mass
of cold gas produces a higher star formation rate in a high-redshift
galaxy.

\item {\it Collisional Starburst (CSB)} : quiescent star formation is
modeled using the ``constant efficiency'' recipe, and in addition,
following mergers, a burst mode of star formation is included using
the recipe described above.
\end{enumerate}
These three models produce similar galaxy properties at low redshift,
but differ dramatically at high redshift (see SPF).

In this paper, we choose to normalize the number density of objects in
each model using an adjustable dust parameter.  As in SPF, we assume
that the face-on optical depth of the disk depends on the intrinsic
rest-UV luminosity of the galaxy via: 
\begin{equation}
\tau_{UV} = \tau_{UV,*}\left (\frac{L_{UV,i}}{L_{UV,*}}\right )^{\beta}. 
\end{equation}
This form was suggested as an empirical description of extinction in
low-redshift galaxies by \citet{wh:96}. The actual extinction is then
calculated by assigning a random inclination to each galaxy and using
a ``slab'' model (see SPF for details). As shown in SPF, this very
simple recipe gives remarkably good agreement with the distribution of
extinctions for observed LBGs derived by \citet{as:00} based on the
slope of the UV continuum.  The parameter $L_{UV, *}$ is taken to be
the observed value of $L_{*}$ given by \cite{steidel:99}, and we fix
$\beta = 0.3$, since this value results in the best fit to the
luminosity function and is still consistent with the results of
\citet{wh:96}.  The value of $\tau_{UV,*}$ is then adjusted separately
for each of the three models in order to match the observed number
density of LBGs.  Given the assumptions we make for the selection
function and number density, the values of $\tau_{UV,*}$ obtained are
0.35, 2.1, and 2.65, for the CEQ, CSB, and AQ model, respectively. A
value of $\tau_{UV, *}=1.75$ would correspond to an extinction
correction of a factor of five, the average value assumed by
\citet{steidel:99}. The more recent results of \citet{as:00} suggest
an average extinction of a factor of $\sim 7$, corresponding to
$\tau_{UV, *}=2.1$. The extinction required by the CEQ model is
therefore a bit low, and for the AQ model a bit high, compared to the
best current observational estimates. As these estimates are still
fairly uncertain, however, this is not a very serious concern. Note
that the number density obtained in the models could be adjusted by
tuning other parameters, but at the possible expense of agreement with
other data.

\subsection{Halo Occupation Functions}
\label{sec:sims:occ}
\begin{figure*}
\resizebox{0.47\textwidth}{!}{\includegraphics{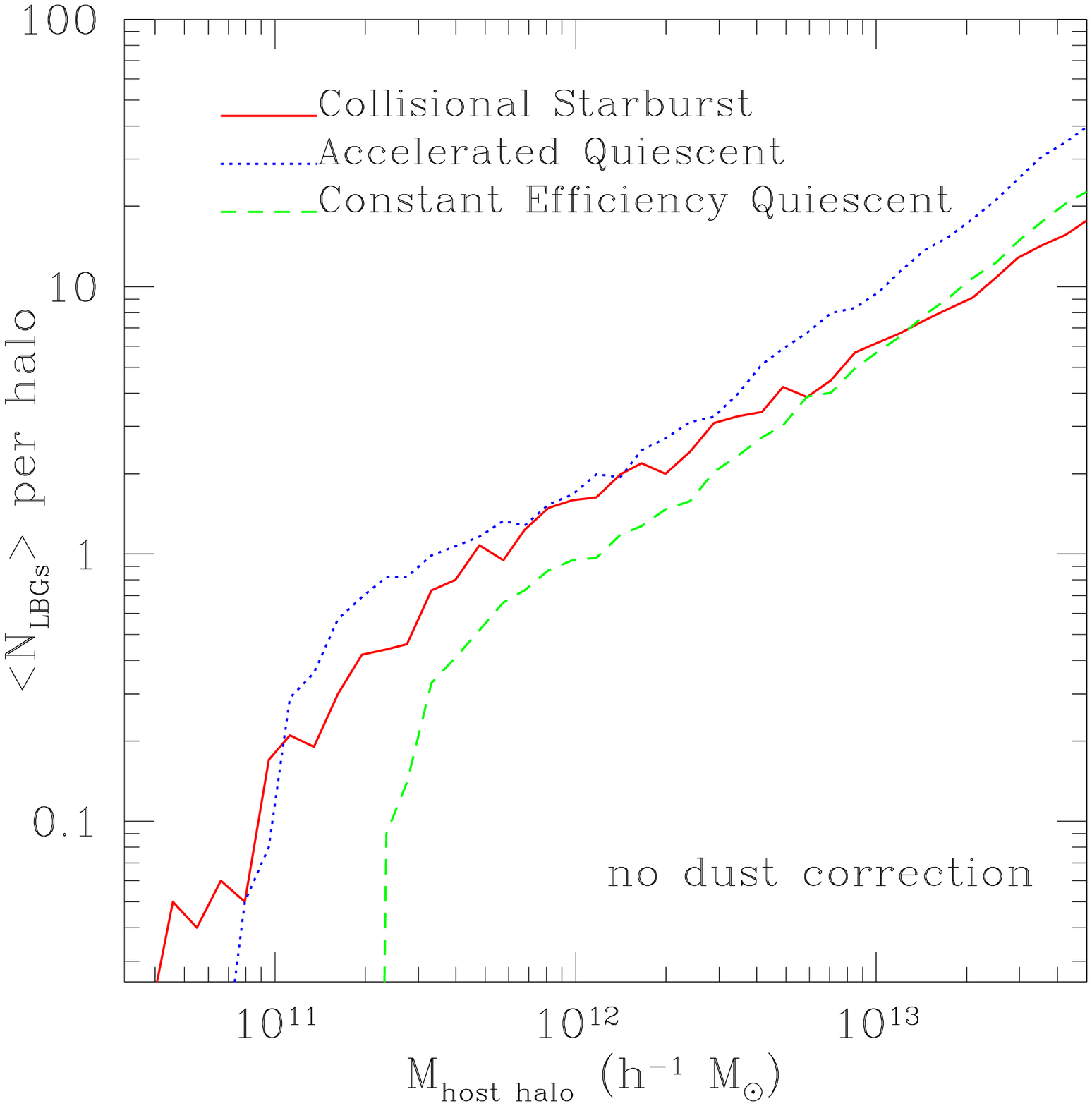}}
\resizebox{0.47\textwidth}{!}{\includegraphics{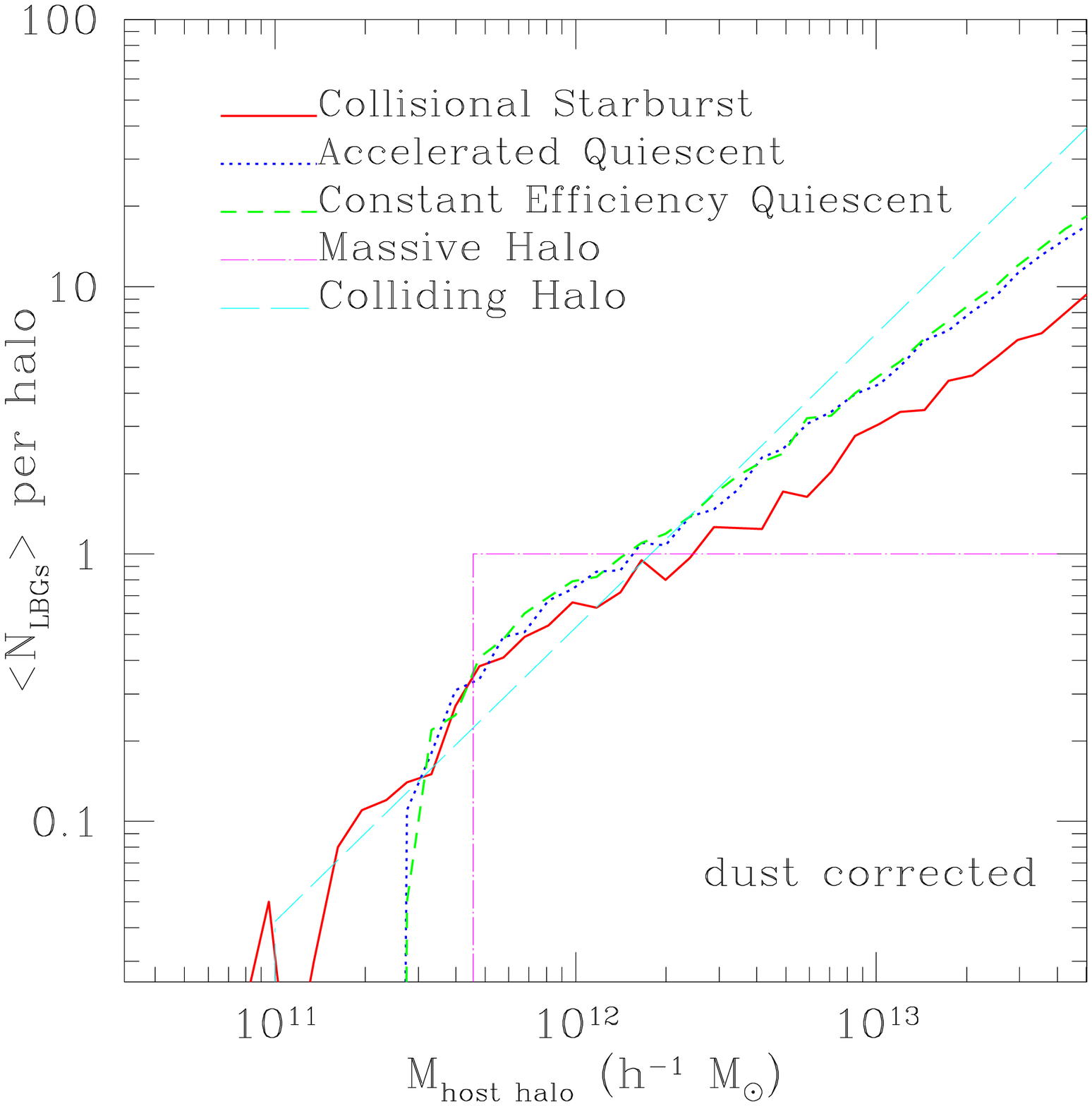}}
\caption{(Left) Average number of galaxies with ${\cal R} \leq 25.5$
per halo for the three semi-analytic models, before (left) and after
(right) dust has been added. On the right-hand panel, the occupation
functions for the Massive Halo and Colliding Halo models
are shown for comparison.
\label{fig:lbgprob}}
\end{figure*}

The semi-analytic model tells us the probability of observing a galaxy
of a given luminosity in a host halo of a given mass.  Specifically,
we take from each model the probability of observing $N$ objects
brighter than ${\cal R}=25.5$ in a halo of mass $M$. In practice, we
run a grid of 50 halo masses, and produce 100 Monte Carlo realizations of
each mass. Figure \ref{fig:lbgprob} shows the average number of
objects per halo with ${\cal R}<25.5$ as a function of mass for each
model, both before and after dust has been added using the approach
described above. The occupation functions for the Massive Halo and
Colliding Halo models are also shown.

The first thing to note is that all of the models, including the
massive quiescent type (represented here by the CEQ model) have
much steeper occupation functions than the Massive Halo
model. This implies that multiple galaxies in high-mass halos are
important even for this class of models. In fact, after the
re-normalization using the dust parameter, the quiescent models
actually have more multiple galaxies in the massive halos than the
Collisional Starburst model.
A power-law functional form similar to the one considered in \S
\ref{sec:mw-lc} provides a good description of all of
the semi-analytic models, with $N_{LBG} \propto M^{0.8}$ on scales
larger than a few $\times 10^{11} \hMsun$ for the two quiescent models, and a
slightly shallower slope of 0.7 for the CSB model.

It is also interesting that the slope of the occupation function for
the semi-analytic Collisional Starburst model, $S \simeq 0.7$, is so
much shallower than that for the Colliding Halo model, $S \simeq 1.1$.
This must be either because the approximations used to model the
collisions of halos in the semi-analytic models are inaccurate, or
because of the more detailed modeling of the luminosity associated
with each collision in the semi-analytics.  This is investigated in
detail in \S\ref{sec:compare}; it is primarily due to the luminosity
assignment, in the sense that mergers are less likely to produce
visible LBGs in massive halos.

Figure~\ref{fig:lbgprob} shows only the \emph{mean} number of galaxies
in halos as a function of their mass; an important additional piece of
information is the scatter in this quantity. \citet{benson:00} have
shown that the scatter is important in determining the small-scale
clustering properties. For the Massive Halo model, we simply assume
that each halo has zero or one galaxy, with no scatter. For the
Colliding Halo model, the number of galaxies is drawn from a Poisson
distribution.  For the semi-analytic models, the scatter is provided
from 100 Monte Carlo realizations of each halo.

Once the number of galaxies is chosen, they must be assigned positions
within the halo; the first galaxy is placed at the center of the halo,
and the additional galaxies are placed randomly in radius within
$R_{\rm vir}$, which corresponds to an isothermal density distribution,
and is also in rough agreement with the results of the ART simulation.
This placement is somewhat uncertain; however, none of the statistics
considered here are very sensitive to the internal structure of the
halo.

\section{COMPARING MODELS WITH DATA: RESULTS}
\label{sec:results}
\subsection{Weighted Overdensity}
\label{sec:results:od}
A standard statistic for measuring the clustering of a population is
the overdensity in some region; in \citet{wech:98}, we looked at the
distribution of LBG overdensities in cells that were $\Delta z = 0.04$
in redshift and $9\arcmin \times 9\arcmin$ on the sky, and compared to
the data from just one field --- 13 cells (from A98).  Explicitly, 
the raw counts $N_i$ were de-selected into
\begin{equation}
{\cal N}_i= N_i/S_i ,
\label{eq:deselect}
\end{equation}
where $S_i$ is the selection function in pixel $i$. From then on, using the
statistic 
\begin{equation}
d_i \equiv \delta{\cal N}_i/\bar{\cal N} = \delta N_i/\bar N_i , 
\label{eq:d}
\end{equation}
where $\delta N = N/\bar{N} -1$, 
all the pixels were treated equally. By doing that, we ignored the fact
that the Poisson errors, which depend on $S_i$, affect the probability
distribution function (PDF) of $d_i$.  In particular, it is ``easier''
to obtain more extreme density contrasts where the error in that
quantity is larger, i.e., where $S_i$ is smaller.  This rather gross
approximation was worst when assigning a single value of $p$ (the
probability of getting a spike of a particular size in one pixel) to
all the pixels and then translating it to $P_1$ (the probability that
a spike of this size is chosen in all pixels); in fact, the actual
probability $p_i$ should vary with $S_i$, and $P_1$ should be computed
accordingly.  Our excuse, which was fine as a first approximation, was
that only pixels with $S_i \geq 0.4 S_{max}$ were included, and thus
the error was kept relatively small.

With the extended data from several fields we can now be more
accurate, and can also include pixels with smaller $S_i$.  The first
goal is to find a statistic that would indeed put all the pixels on
the same footing. Such a statistic is the error weighted galaxy
overdensity: \beq D_i \equiv {\delta {\cal N}_i/\bar{\cal N}
\over \sigma_i}, \eeq where $\sigma_i$ is the Poisson error in the
quantity of interest, $d_i$, which measures fluctuations in the real
universe.

Since the Poisson error in $N_i$ is $\bar N_i^{1/2}$ (ignoring additional
factors proportional to $J_3$ in case of correlations), it follows from
equations~\ref{eq:deselect} and \ref{eq:d} that $\sigma_i= \bar N_i ^{-1/2}$,
and thus  
\begin{equation}
D_i = {\delta N_i \over \bar N_i ^{1/2}}.
\end{equation}
The square-root $\bar N_i ^{1/2}$ in the denominator replaces the
$\bar N_i$ in the denominator of the old statistic. With the new
statistic, a spike of a given positive true relative over-density $d_i$
that occurs where $S_i$ is lower than $S_{max}$ is now associated with
a smaller $D_i$ compared to a spike of a similar $d_i$ at
$S_{max}$. This takes into account the fact that larger density
contrasts are more likely to occur where $S_i$ is small.

The statistic $D_i$ describes the count fluctuations in terms of the
rms Poisson fluctuations in each pixel.  If there are enough LBGs
expected on the average, the distribution of $D_i$ approaches that of
a Gaussian with width unity.  Then it is perfectly consistent to
consider all the pixels together on the same footing, and one could
use Gaussian statistics to evaluate probabilities.  Since we are not
really in the Gaussian limit, the PDFs in the different pixels are not
exactly the same, although they are far closer to each other than
before.  To deal with this imperfection, the comparison of the data
and models is pursued in the ``observational plane'': we apply the
observed selection function to the simulated counts and then compute
the statistic $D_i$ and construct its PDF by the distribution of its
value over the pixels. This PDF is then compared to the PDF
constructed directly from the data.

\begin{inlinefigure}
\begin{center}
\resizebox{\textwidth}{!}{\includegraphics{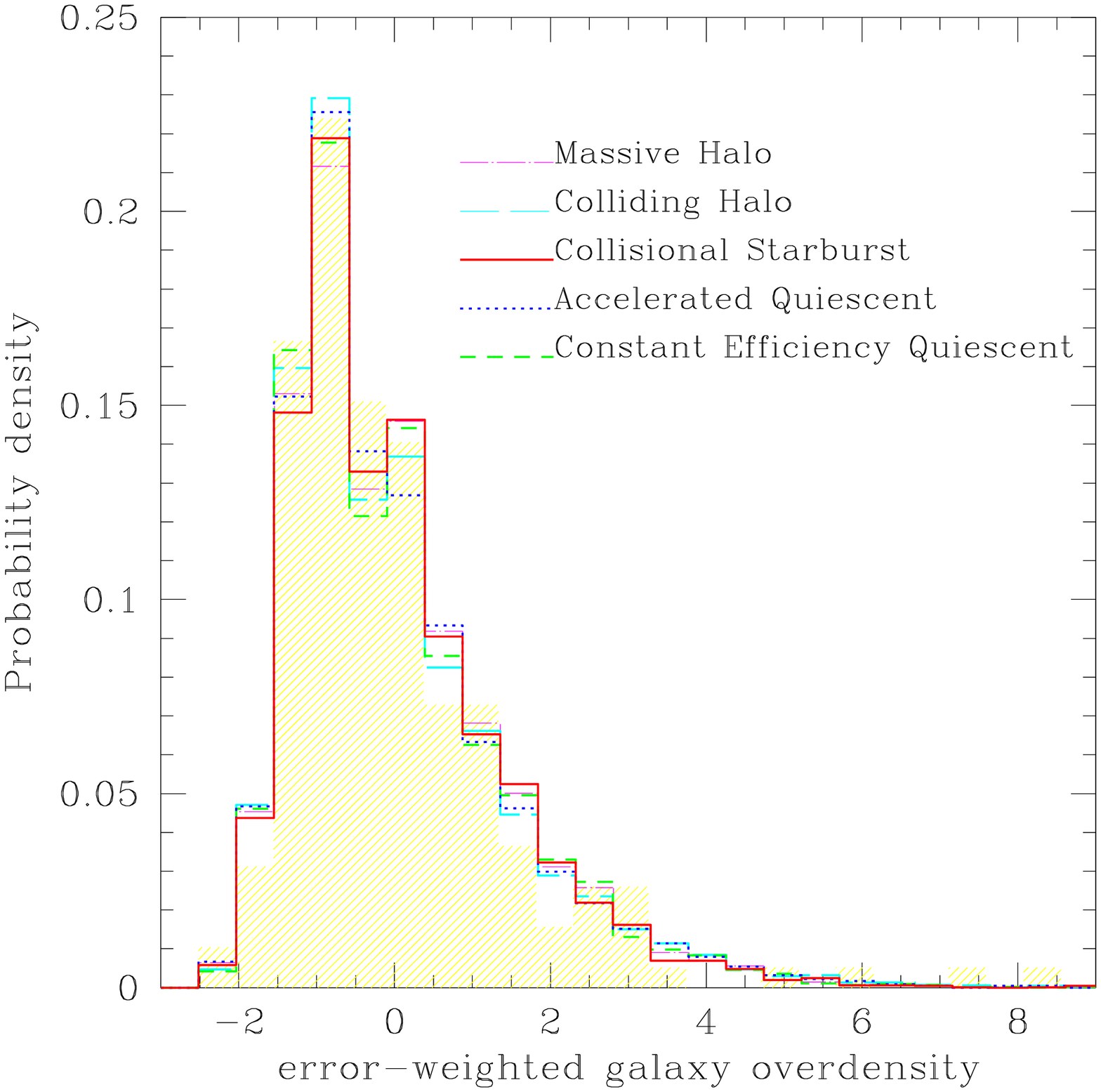}}
\end{center}
\figcaption{Probability distribution of the
error-weighted galaxy overdensity for the five models, compared with
eight fields of data (shaded), from A98 and \citet{adel:01}. \label{fig:d}}
\end{inlinefigure}

For each model, the differential distribution of this statistic is
compared with that of the data (Figure \ref{fig:d}) using the
Kolmogorov-Smirnov (KS) test, which gives the probabilities that the
data and the model came from the same underlying distribution. The
results are shown in Table \ref{tab:results}, and show that none of
the models can be ruled out.  The KS statistic, however, can
systematically underestimate the significance of differences between
the observations and the models, especially if the differences are
near the ends of the distribution \citep{press:92}. Kuiper's variant
of this test \citep{kuiper:62,press:92} uses the sum of the maximum
positive difference and the absolute value of the maximum negative
difference, instead of the the maximum of the absolute value of the
difference between observed and expected cumulative counts used by the
standard KS test, and does not suffer from these problems.  The values
for this test are also listed in Table \ref{tab:results}.  In this
analysis, there are 192 data pixels and 720 simulation pixels, each
$9\arcmin \times 9 \arcmin$ on the sky and $\Delta z=0.04$ in
redshift.  The assignment of galaxies to halos and ``observation'' of
LBGs is done 10 times for each model; the numbers quoted in the table
are the mean and error on the mean of these runs.  Unfortunately, none
of the models can be ruled out even using this modified statistic;
even the two extreme halo models cannot be distinguished from the data
at present.  However, it should be noted that we are comparing to an
observational sample with only 500 galaxies.  With two to three times
more data (much of which already exists but is unpublished), these
statistics will become discriminatory.
\begin{table*}
\caption{Models of Lyman-Break Galaxies}
\begin{center}
\begin{tabular}{lcccc}
\tableline
\tableline	
Model & halo occupation: $N(M)\propto M^S$ & normalization & star formation/luminosity assignment\\ 
\tableline
Massive Halo (MH) & $S=0$, no scatter & Mass cut & 
$L \propto M$, $M >4.5 \times 10^{11}$\\
Colliding Halo (CH)	& $S=1.1$, Poisson scatter & C, $N(M)=CM^S$ & $L\propto(M_1+M_2), M_{\rm host}>10^{11}$\\
Collisional 		& semi-analytic, $S \sim 0.7$ 
& dust, $\tau_{UV}= 2.1$& quiescent, $\tau_* = $ constant\\
Starburst (CSB) &&& + starbursts\\
Constant Efficiency	&  semi-analytic, $S \sim 0.8$ 	
& dust, $\tau_{UV}= 0.35$& quiescent, $\tau_*$ = constant\\
Quiescent (CEQ) 		& & & \\
Accelerated 		&semi-analytic, $S \sim 0.8$ 	
& dust, $\tau_{UV}= 2.7$& quiescent, $\tau_* \propto t_{\rm dyn}$\\
Quiescent (AQ) 		& & & \\
\tableline
\end{tabular}
\tablecomments{Model parameters for populating halos with visible galaxies: 
the halo occupation, the method by which the model is normalized to
match the number density of observed objects, and the method of assigning
luminosities to determine which galaxies are visible.}
\end{center}
\label{tab:models}
\end{table*}

\begin{table*}
\caption{Kolmogorov-Smirnov and Kuiper Probabilities and 
	Correlation Function Parameters}
\begin{center}
\begin{tabular}{cccccccc}
 \multicolumn{1}{l}{} &
 \multicolumn{2}{c}{PDF probabilities} &
 \multicolumn{2}{c}{Counts-in-Cells} &
 \multicolumn{3}{c}{3-space Correlation Function} \\
\tableline
\tableline
Model& K-S & Kuiper& $\sigma^2_{gal}$&$ r_0$ [\hmpc]&
$r_0$ [\hmpc]& $r_0$ [\hmpc] & $\gamma$\\
&probability& probability & & $\gamma$=1.6&$\gamma$=1.6 & $\gamma$ free & \\
\tableline
MH	& $0.71 \pm 0.06$ & $0.66 \pm 0.08$ &  $0.69 \pm 0.10$ & $4.1 \pm0.4$ &$4.62 \pm 0.18$&$ 4.65 \pm 0.20$&
$1.51 \pm 0.09$\\

CH 	& $0.92 \pm 0.03$ & $0.82 \pm 0.03$ & $1.19 \pm 0.18$& $5.9 \pm 0.5$ &$5.59 \pm 0.27$&$ 5.60 \pm 0.30$&
$1.62 \pm 0.10$\\

CSB	& $0.73 \pm 0.05$ & $0.78 \pm 0.06$ & $0.69 \pm 0.11$ & $4.2 \pm 0.5$ &$4.62 \pm 0.24$&$ 4.64\pm 0.24$&
$1.52 \pm 0.11$\\

CEQ	& $0.81 \pm 0.06$ & $0.76 \pm 0.04$ & $0.90 \pm 0.13$&$4.9 \pm 0.4$ &$5.26 \pm 0.21$&$5.29\pm 0.24$&    $1.58 \pm 0.09$\\

AQ	& $0.80 \pm 0.04$ & $0.71 \pm 0.06$ & $0.88 \pm 0.12$ &$4.9 \pm 0.4$ &$5.16 \pm 0.28$&$5.20 \pm 0.29$&
$1.57 \pm 0.10$\\
\tableline
\end{tabular}
\tablecomments{The Kolmogorov-Smirnov and Kuiper probabilities that
the overdensity distribution of each model is consistent with the
data.  Also listed is the variance ($\sigma^2_{gal}$) in counts in
cells of 11.4 \hmpc, and the correlation length derived from this
value, for galaxies in each model brighter than ${\cal R}_{AB}=25.5$.
For each model, we also list the best fit correlation length $r_{0}$
for fixed slope $\gamma=1.6$, and the best fit values for $r_{0}$ and
$\gamma$ when fit independently. All fits to the correlation function
are performed over the range 1 \hmpc\ $ \leq r \leq 8$ \hmpc.}
\end{center}
\label{tab:results}
\end{table*}

\begin{table*}
\caption{Observational Correlation Function Parameters}
\begin{center}
\begin{tabular}{cccccc}
\tableline
\tableline
Sample & Method & magnitude limit & $r_0$ [\hmpc] & $\gamma$ & reference \\
\tableline
SPEC  & CIC & ${\cal R}=25.5$ & $6 \pm 1$ & [1.8] & Adelberger et al. 1998 \\
SPEC & CIC & ${\cal R}=25.5$ & $4.4 \pm 0.9$ &  $[1.6]$ & Adelberger 2000 \\
SPEC & $w(\theta)$ & ${\cal R}=25.5$ & $3.8 \pm 0.3$ & $1.61 \pm 0.15$ & Adelberger 2000\\
SPEC & CIC & ${\cal R}=25.0$ & $5.0 \pm 0.7$ &  $[2.0]$ & Giavalisco et al. 2001 \\
PHOT & $w(\theta)$ & ${\cal R}=25.5$ & $3.2 \pm 0.7$ & $2.0\pm 0.2$ & Giavalisco et
al. 2001\\
HDF & $w(\theta)$ & $V_{606}=27$ & $1.2^{+0.9}_{-0.8}$ & $2.2^{+0.6}_{-0.3}$ & Giavalisco et
al. 2001\\
HDF photo-z & $w(\theta)$ & $I_{814}=28.5$ & $2.78 \pm 0.68$ & [1.8]
& Arnouts et al. 1999\\
\tableline
\end{tabular}
\tablecomments{The correlation function parameters derived from the
observations, for several different samples and methods, assuming the
same \LCDM\ cosmology used throughout our analysis. SPEC refers to the
ground-based spectroscopic sample, PHOT to the ground-based sample of
photometric LBG candidates, and HDF to the deeper sample of $U_{300}$
drop-outs from the HDF North. HDF photo-z is the sample of HDF
galaxies with photometric redshifts in the range $2.5 < z < 3.5$. All
magnitude limits are given in the AB system, and are the authors'
stated completeness limits (note that the SPEC samples of
\protect\citealt{adel:98} and \protect\citealt{giav:00} are just subsamples of the \protect\citealt{adel:99} sample).  CIC
refers to the counts-in-cells method and $w(\theta)$ to the inversion
of the angular correlation function. Where $\gamma$ is given in square
brackets, this indicates that the value was assumed rather than
derived. }
\end{center}
\label{tab:cf}
\end{table*}

\subsection{Two-Point Correlation Function}
\label{sec:results:cf}
We use the standard notation for that ubiquitous measure of
clustering, the correlation function, $\xi(r)=(r/r_{0})^{-\gamma}$.
The observed correlation length $r_0$ of a sample with redshift
information may be estimated either from a counts-in-cells analysis
(assuming a value for $\gamma$), or by inverting the angular
correlation function (e.g., \citealt{peebles:80}).  For our usual
\LCDM\ cosmology, the initial estimate using the first method yielded
a value of $r_{0}
\sim 6$ \hmpc\ (A98), whereas the second method yielded lower
values of $\sim 3-4$ \hmpc\ \citep{adel:99}.  However, more recent
observational estimates give lower values of roughly $r_{0} \sim 4.4$
\hmpc\ and $r_{0} \sim 3.1$ \hmpc\ (K. Adelberger 2000, private communication)
for the counts-in-cells and $w(\theta)$ methods respectively.  

Using the angular correlation inversion method on a fainter sample
($V_{606} < 27$) of LBGs in the HDF, G00 obtained even smaller
correlation lengths, $1.4-1.7$ \hmpc. However, the analysis of
\cite{arnouts:99}, based on galaxies in the HDF with photometric
redshifts in the range $2.5 < z < 3.5$ and $I_{814}<28.5$, yields
$r_0\sim 3$ \hmpc, consistent with the brighter ground-based samples
(see also \citealt{mag:99}). The correlation function parameters
obtained from the observations, calculated for our \LCDM\ cosmology,
are summarized in Table~\ref{tab:cf}. We return to the possibility of
luminosity segregation in \S\ref{sec:results:lum}, and for the moment
concentrate on the brighter (${\cal R}_{AB} < 25.5$) ground-based samples 
with spectroscopic redshifts.

Since different methods of estimating the correlation length may give
different values, and since the selection function of the
observational sample may also affect the result, we calculate the
correlation length from our simulations in two ways.  First, we simply
calculate the real-space correlation function in three dimensions,
using all of the galaxies brighter than ${\cal R}_{AB} < 25.5$ in
each model, randomly sampled to match the observed number density
(different selection probabilities for different regions are not used
in selecting galaxies for this method, since this would bias the
results).  The real-space correlation function for all five of our
models is shown in Figure \ref{fig:cf}.  The errors quoted represent
the $1\sigma$ scatter in the results of 100 resamplings and
reassignments of galaxies to halos.  The best fit values for the
correlation length are listed in Table \ref{tab:results} for $\gamma$
fixed to 1.6 and for $\gamma$ left as a free parameter.  In each case
we only fit the data on scales between 1--8 \hmpc, where the errors
and the deviation from a power law are small; we concentrate on scales
smaller than this in the next section.

The counts-in-cells method estimates the correlation length by
measuring the variance of galaxy counts in spatial bins of a given
size:
\begin{equation}
\sigma^2_{\rm gal} = [\langle (N-\mu)^2\rangle-\mu]/\mu^2,
\end{equation}
where $\mu$ is the expected number of galaxies in a cell, equal to
the total number density of observed galaxies times the probability of
observing a galaxy in that cell. Subtracting the $\mu$ term removes
shot noise, since the average number of galaxies per cell is small.
We follow the method of \citet{adel:98} as closely as possible to
estimate this statistic for our sample: we break the box into cubical
cells, which for this cosmology have a length of 11.4 \hmpc, select
the galaxies in each cell with a fraction drawn from one of the data
cells, calculate the estimator $[(N-\mu)^2-\mu]/\mu^2$ for each cell,
and then combine the estimates from each cell with inverse-variance
weighting for a final estimate of $\sigma^2_{\rm gal}$.  If the
correlation function is a pure power law, for spherical cells the
correlation length is given by: $r_0 = R_{\rm cell}[\sigma^2_{\rm
gal}(3-\gamma)(4-\gamma)(6-\gamma)2^{\gamma}/72]^{1/\gamma}$
\citep{peebles:80}.  The values of $\sigma^2_{\rm gal}$, and the
corresponding values of $r_0$ (taking $R_{\rm cell}$ to be the radius
of a sphere with volume equal to that of our cubical cells) are given
in Table \ref{tab:results}.  These should be compared to the current
value obtained from the observational sample: $\sigma^2_{\rm gal} =0.75
\pm 0.25$ (\citealt{adel:99}; note that the value
from the earlier published work of A98 was  $\sigma^2_{\rm gal}=1.3
\pm 0.4$).  If, instead of the selection procedure described above, 
each cell is just randomly selected with the same probability, the
results are essentially unchanged.  Note that the errors listed in the
table are the variance over 100 re-samplings of our entire box; the
variance over regions the size of the full data sample used here
(approximately three times smaller) is quite close to the error
quoted on the data --- roughly ~0.2 in $\sigma^2_{\rm gal}$.
Unfortunately, it is not straightforward to calculate the angular
correlation function from the simulation in a way that would be
meaningful for comparison to the data, since our box is not large
enough to have the same angular projection effects as the data.

\begin{inlinefigure}
\begin{center}
\resizebox{\textwidth}{!}{\includegraphics{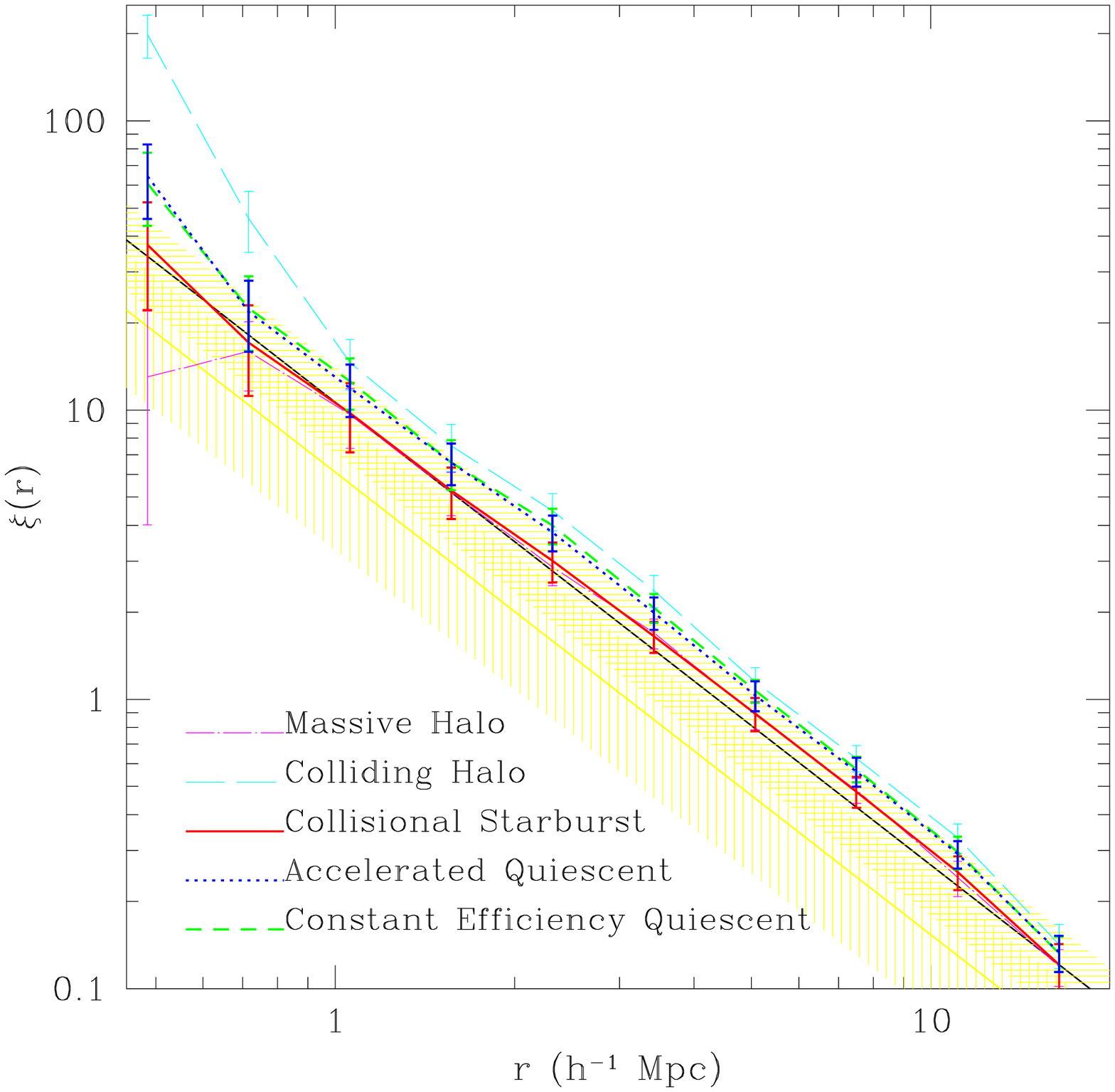}}
\end{center}
\figcaption{Correlation function for all five models.
Also plotted are the most recent best-fit parameters, with shaded
error regions, from the observations, for the counts-in-cells method
(horizontal shading), and the inversion of the angular correlation
function (vertical shading).
\label{fig:cf}}
\end{inlinefigure}

The two methods presented here give fairly similar results, although
for most of the models the counts-in-cells method gives a slightly
lower value than that estimated directly from the three-space
correlation function.  The biggest discrepancies are for the Massive
Halo and Colliding Halo model: in the former the counts-in-cells
method gives a significantly lower value, in the latter it gives a
higher value.  The reason for this is easy to understand --- the
counts-in-cells method is sensitive to clustering on all scales
smaller than the cell size, and assumes that the correlation function
is a power law over this full range.  As can be seen from
Figure~\ref{fig:cf}, in the Massive Halo model, the correlation
function is shallower than a power law on small scales, while in the
Colliding Halo model, it is steeper.

The various models actually have quite similar correlation lengths,
and all of the models, with the possible exception of the Colliding
Halo model, are within reasonable agreement with the counts-in-cells
estimate from the data.  The latest estimate, for the same sample,
from the inversion of the angular correlation function, however, is
quite a bit lower (see Table~\ref{tab:cf}) --- if this value turns out
to be correct, all of the models presented here may be in trouble.
There may be more hope of distinguishing the models using their
clustering on small scales; we focus on this in the following section.

\subsection{Close Pairs}
\label{sec:results:pairs}
From examining Figure~\ref{fig:lbgprob}, it is clear that a major
difference between our five models is the number of multiple objects
within one halo.  Although this cannot be directly observed, one can
determine the number of pairs of objects at small angular separations
in the models, and compare directly to observations.  ``Pairs'', in the
sense used here, are objects within a given angular separation which are
also within a redshift interval of $\Delta z = 0.04$.  This definition
is used for both the data and the models.

Figure~\ref{fig:pairs} shows, for angular separations between 0 and
$60\arcsec$, the number of pairs divided by the total number of
galaxies for all five models, compared with the data.  One might be
concerned that the true number of close pairs would be underestimated
if there was a bias against obtaining spectra for close pairs, due for
example to the physical limitations of slit placement on the masks.
However, each field is typically observed with several independent
masks so that this effect is not very large.  For example, for a
sub-sample of candidates that includes 109 pairs of objects within $10
\arcsec$ of each other, spectroscopy is obtained for half of the
objects, and is obtained for both objects in 21 of the pairs
(K. Adelberger 1999, private communication), instead of the number
that would be expected with no bias --- $109\times 0.5^2 = 27.25 \pm
5.2$.  This suggests that the systematic error from selection against
close pairs is less than about 25\%.

In Figure \ref{fig:pairs}, the only significant differences between
the models are in the first bin, which at $z\sim 3$ corresponds to a
comoving distance of $\sim 300$ \hkpc\ (a physical size of $\sim
80$\hkpc, which is roughly the virial radius of a $10^{12}\msun$ halo)
for this \LCDM\ cosmology, and includes most galaxies that are in the
same halo.  However, models which are dominated by galaxies in more
massive halos will depend more sensitively on the distribution within
the halo.  Still, the determining factor in the number of pairs in
this bin is mainly the number of multiple galaxies in massive halos,
or, since all the models have been normalized to have the same total
number density, the slope and cutoff of the halo occupation function,
$N_{LBG}(M)$, that was discussed in \S \ref{sec:sims:occ}.  
The Massive Halo model, in which all the pairs reside in
different halos, underpredicts the number density of close pairs by
more than $1.5 \sigma$.  Conversely, the Colliding Halo model
overpredicts the number of pairs within $15 \arcsec$ by almost $4
\sigma$.  However, we find that all three ``realistic'' semi-analytic
models, including the Collisional Starburst model, match the
data at least reasonably well, especially given that the number of
spectroscopic pairs may be underestimated somewhat.  In fact, the
quiescent models actually predict more close pairs than the
Collisional Starburst model.  This is counter to na\"{\i}ve
expectations, but follows from what we found in \S \ref{sec:sims:sams}
--- that each massive halo actually has more LBGs in the quiescent
models.

\begin{inlinefigure}
\resizebox{\textwidth}{!}{\includegraphics{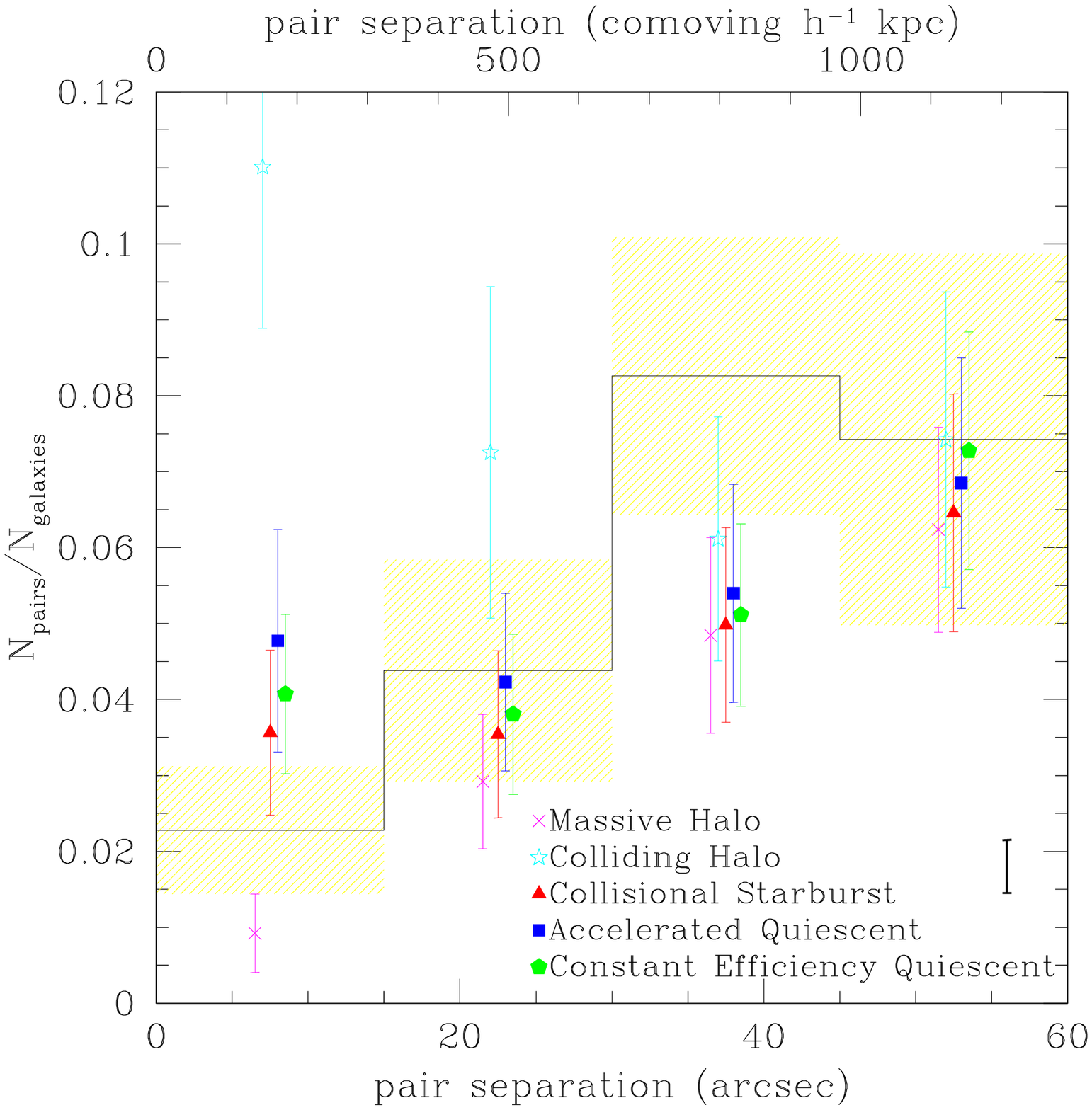}}
\figcaption{Number of pairs divided by total number of galaxies for the models
(symbols; the plotted horizontal locations are shifted slightly for clarity), 
compared with the data ($1\sigma$ error on the mean of eight fields 
is plotted with shaded boxes).  Errors plotted for
the models are the $1\sigma$ scatter of 50 re-samplings each of three
different regions of the box, each the same volume as the total data sample.
The typical error due to cosmic scatter is shown in the lower right 
corner of the plot.
\label{fig:pairs}}
\end{inlinefigure}

Although the models primarily differ from each other on spatial scales
smaller than the size of the most massive halos (which are mostly
within the first and second bin of Figure \ref{fig:pairs}), all of the
models seem to underpredict the number of pairs at intermediate
separations, specifically from 30-45\arcsec --- though the number of
observed objects suffers from small number statistics, and this seems a likely cause of the discrepancy.  At these
separations, the close pair statistics are mostly determined by the
clustering of the dark halos. Adjusting the cosmology or other details
of the models may improve the predictions, or perhaps the selection is
not completely understood.  It should also be noted that we have
ignored any correlation in the galaxy properties on scales larger than
the halos themselves --- i.e., the dependence of galaxy formation
efficiency on the large-scale environment (obviously, correlation
between the dark halos themselves is built into the halo catalog). One
might imagine, for example, that the details of the galaxy
luminosities could be dependent on the merger history of the halos,
which may depend on the larger scale environment. This is explored
further in future work \citep{wech:01}.

As noted earlier, and as emphasized by \citet{benson:00}, the
clustering strength on small scales depends on the scatter in the halo
occupation function.  In the models presented here, we assumed no
scatter for the Massive Halo model, Poisson scatter for the Colliding
Halo model, and the actual scatter given by the full semi-analytic
treatment for the other three models.  In all cases, we find that
including scatter increases the small-scale correlations.  The scatter
from the semi-analytic models results in slightly lower correlations
than Poisson.  In the case of the Colliding Halo model, using the mean
decreases the pair fraction in the first bin by about 0.03 --- far
from sufficient to reconcile it with the data.  An alternative to
looking directly at close pairs would be to compare the scale
dependence of the bias.  It is clear from Figure ~\ref{fig:cf} that
the Colliding Halo model is significantly more biased on small scales
than on large scales, and the reverse is true for the Massive Halo
model --- so this might be another possible discriminant if it was
well measured in the data.

\subsection{Dependence of Clustering on Luminosity}
\label{sec:results:lum}

There have been suggestions (S98, G00) that the
clustering strength of LBGs depends on the magnitude limit of the
sample, or similarly on the number density of the population. These
authors compared the correlation length obtained from the ground-based
spectroscopic and photometric samples, and the much deeper sample of
LBGs identified in the Hubble Deep Field (HDF). They found a monotonic
decrease in the correlation length as the magnitude limit of the
sample grew fainter, suggesting that intrinsically brighter galaxies
are more strongly clustered (see Table~\ref{tab:cf}). This result, if
correct, would provide further constraints on the relationship between
visible galaxies of different luminosities and the dark-matter halos
that host them.

In S98 and G00, the authors interpreted their observational results as
evidence for a tight connection between halo mass and UV-luminosity or
star formation rate. Therefore, one might expect that the luminosity
dependence of clustering would provide a good way to distinguish
between starburst models and quiescent models.  In quiescent models,
the star formation is primarily dependent on the mass of cold gas
available to form stars, and thus one might expect a tight correlation
between halo mass and luminosity.  In burst models, on the other hand,
some correlation is expected, but it should be significantly looser
than that of the quiescent models, since the luminosity is dependent
on the details of the merger.

However, the strength of this observational trend is still rather
uncertain.  In particular, the correlation length obtained from the
same data set seems to be dependent on the method used; when derived
from counts-in-cells, the correlation length for the spectroscopic
sample is larger than when derived by inverting the angular
correlation function. It appears that when the same method is used,
similar results for the correlation length are obtained from the
spectroscopic and photometric samples (see Table~\ref{tab:cf}), the
former of which is a somewhat brighter subsample of the later.  Note,
however, that there has yet to be either an analysis of the data which
compares the two methods for {\em exactly} the same sample, or an
analysis which compares the same method for two samples which differ
{\em only} in their magnitude limit.  The correlation length obtained
by G00 for the much fainter sample of Lyman-break galaxies from the
HDF does seem to be considerably lower than that measured from the
ground-based sample.  However, the filter bands and photometric
criteria used to select LBGs in the HDF are different from the
ground-based sample, resulting in a different redshift distribution,
and the small volume probed by the HDF may cause the correlation
length to be underestimated. Moreover, the analysis of
\citet{arnouts:99}, based on a sample from the HDF with a similar
magnitude limit, but selected via photometric redshifts rather than
the Lyman-break technique, yields a correlation length comparable to
that of the brighter ground-based samples. If anything, this result
should be more accurate than the result obtained by G00 because of the
more accurate knowledge of the redshift distribution of the sample.
We therefore consider the strength of the actual observational trend
to be quite uncertain at this point, but explore the model predictions
in any case.

\begin{inlinefigure}
\begin{center}
\resizebox{\textwidth}{!}{\includegraphics{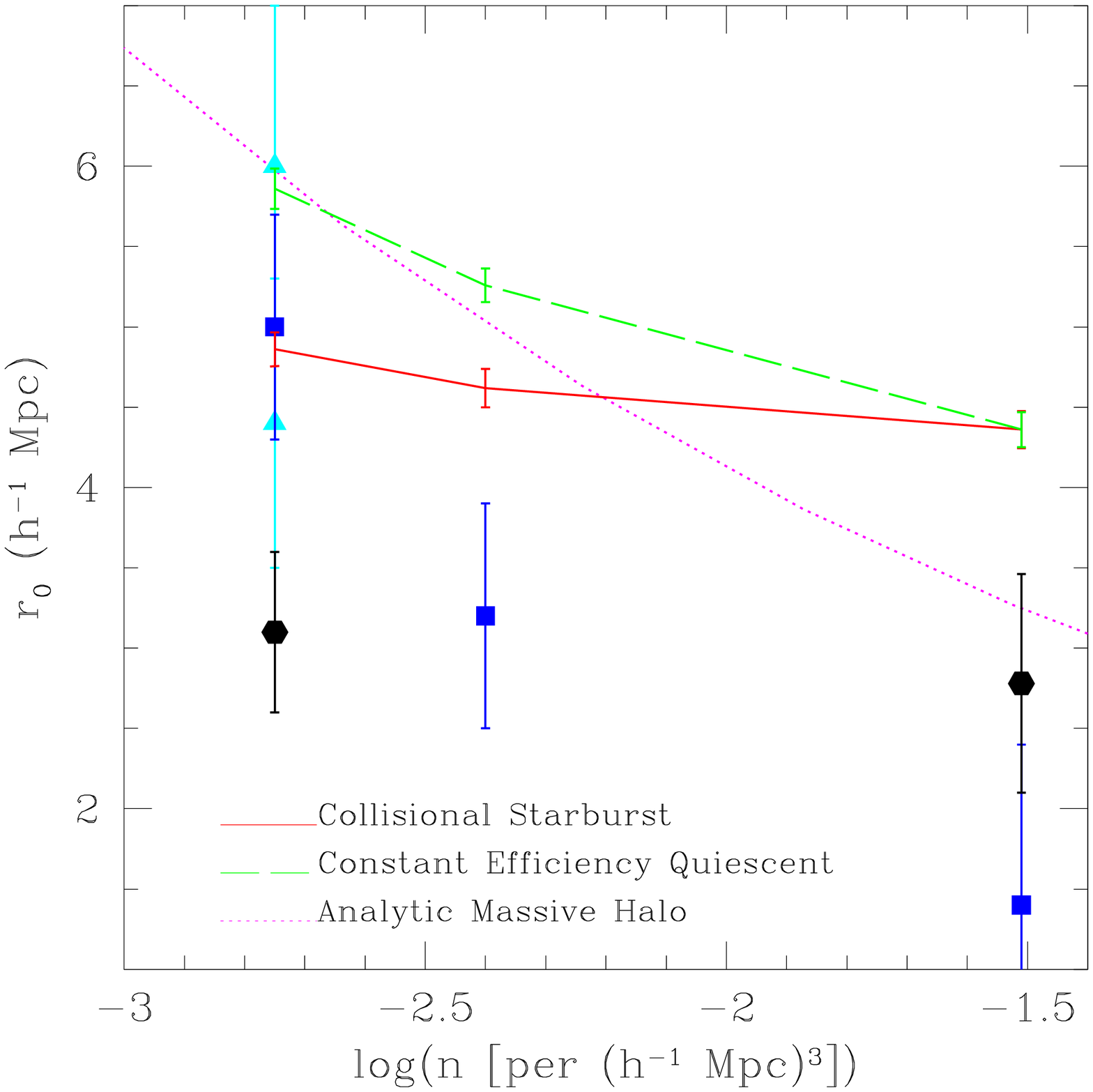}}
\end{center}
\figcaption{Correlation length $r_0$, with fixed $\gamma=1.6$, as a
function of the galaxy number density for two of the semi-analytic
models and the Massive Halo model (calculated using the expression
given by \citealt{jing:99}).  These may be compared to observational
estimates from Adelberger \etal\ using the counts-in-cells method (top
triangle: 1998, bottom triangle: 2000); and using inversion of the
angular correlation function (hexagons), for the sample of Adelberger
\etal\ (left), the HDF sample of
\citet[][right]{arnouts:99}, and the sample of \citet[][squares]
{giav:00}.
\label{fig:r0}}
\end{inlinefigure}

The correlation length is plotted for samples with various number
densities for our Collisional Starburst and Constant Efficiency
Quiescent models in Figure \ref{fig:r0}. There is a weak dependence of
correlation length on number density (magnitude limit) in both models,
especially at the brightest magnitudes, between ${\cal R}_{AB}=25$ and
${\cal R}_{AB}=25.5$.  The trend is slightly stronger in the quiescent
model, as expected.  However, the overall trend in both models is much
weaker than that shown by the analysis of the observations presented
in G00.  The Massive Halo model shows a strong trend, as illustrated
before by S98, \citet{mmw:99}, \citet{arnouts:99}, and G00 (note that
the Massive Halo model in the figure is an analytic model, and not
identical to the results from the simulations --- which don't have
sufficient mass resolution to reach the highest number densities).  If
one ignores the overall offset --- which may be due to systematic
effects coming from the angular correlation function method --- the
trend in the collisional starburst model seems to be the best match to
the most recent data.

\begin{figure*}
\resizebox{0.47\textwidth}{!}{\includegraphics{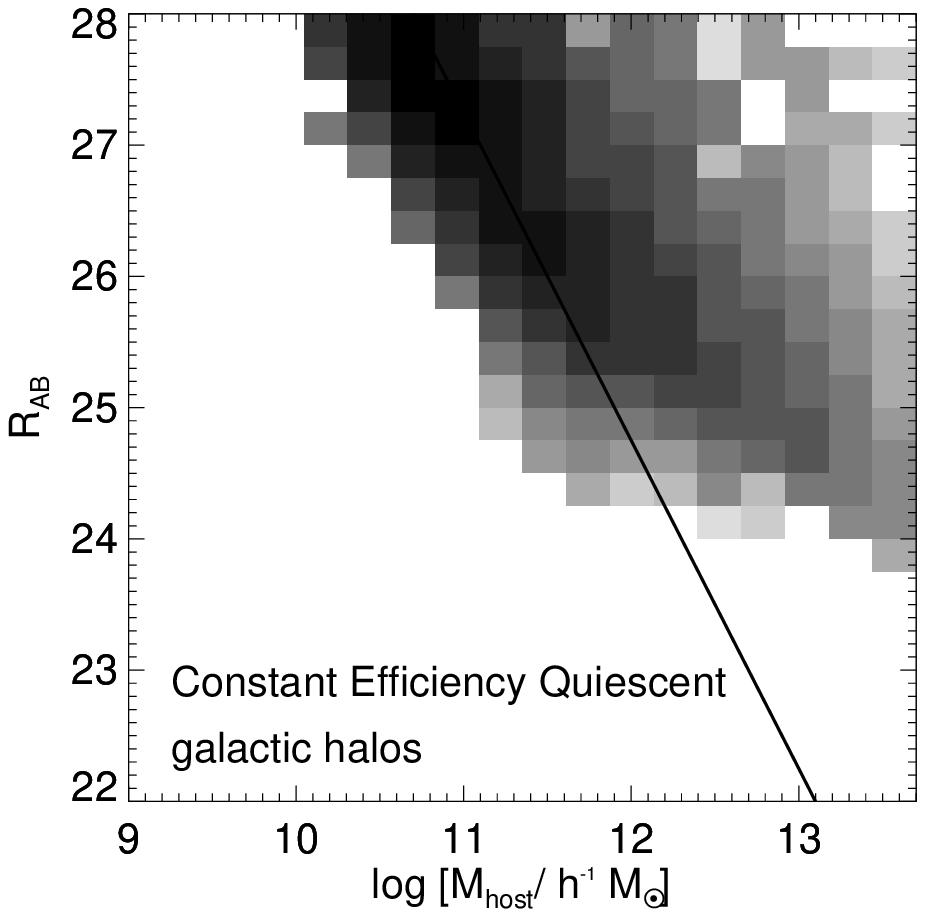}}
\resizebox{0.47\textwidth}{!}{\includegraphics{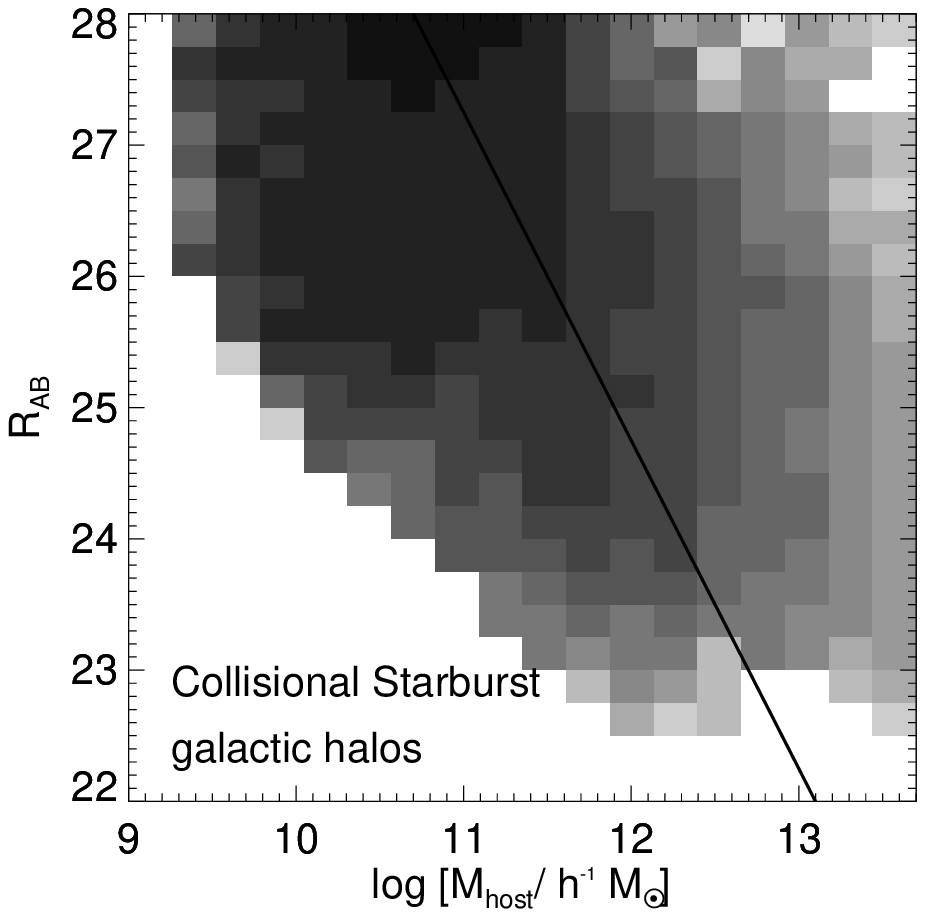}}
\caption{Joint probability distribution of extinction corrected
observed ${\cal R}_{AB}$ magnitude and galactic halo mass (defined
here as the halo the galaxy formed in --- this is usually enclosed
within another, larger halo, but may not be in the case of central
galaxies), for the Constant Efficiency Quiescent (left) and
Collisional Starburst (right) models.  The shadings correspond to
logarithmically spaced density bins, and the line indicates a linear
relation between mass and luminosity.
\label{fig:submags}}
\end{figure*}

To understand why the trend is so weak in the semi-analytic models, we
examine the relationship between galaxy luminosity and halo mass in
Figure~\ref{fig:submags}, in which we show the joint distribution of
observed-frame (rest-UV) ${\cal R}_{AB}$ magnitudes and galactic halo
masses, for both the quiescent and starburst models.  We use the term
``galactic halo'', to refer to the halo that the galaxy directly
resides in.  In most cases this is a subhalo, but in the case of a
central galaxy may be a distinct halo (i.e. not within the virial
radius of a larger halo).  The scatter between galactic halo mass and
galaxy luminosity is smaller in the quiescent model, as expected, but
there is still a significant amount of scatter, resulting from the
differing amounts of cold gas in each galaxy and their different star
formation histories. The luminosity is approximately proportional to
the galactic halo mass for small halos, but for larger halos some of
the cold gas has not yet had time to cool, and the relation departs
from the simple assumption of $L \propto M$. For the starburst models,
$L \propto M$ is a rather poor approximation for all masses, and the
scatter is very large.

\begin{figure*}
\resizebox{0.47\textwidth}{!}{\includegraphics{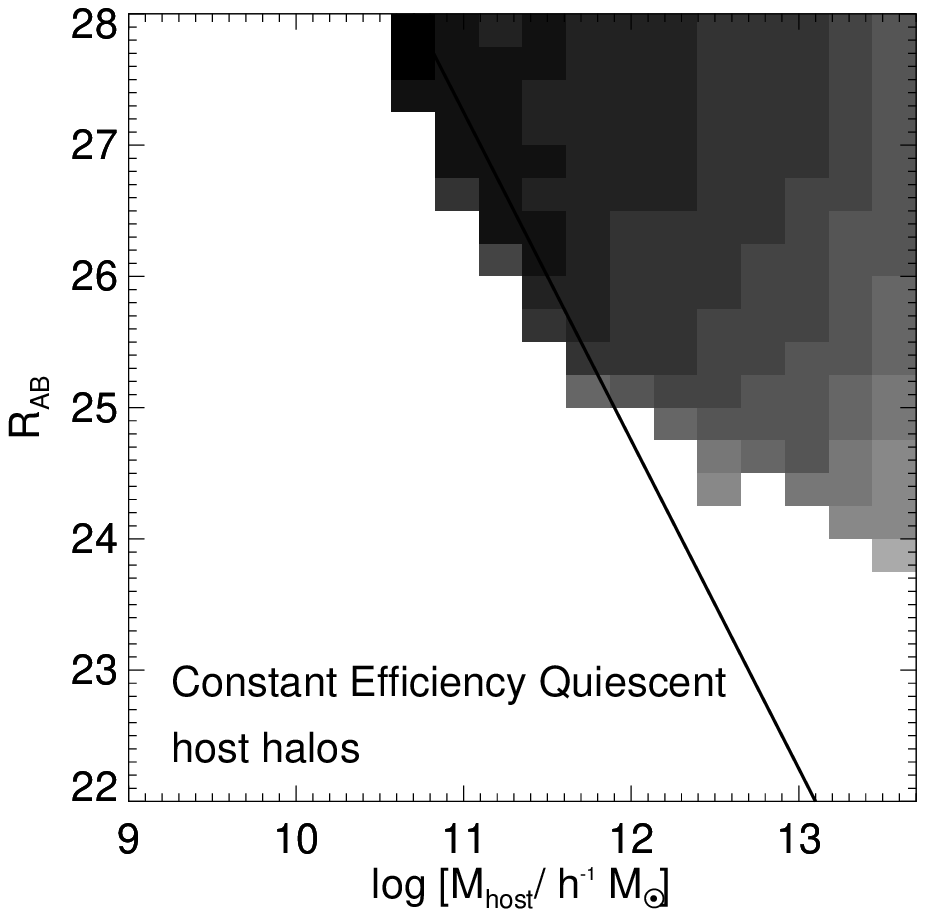}}
\resizebox{0.47\textwidth}{!}{\includegraphics{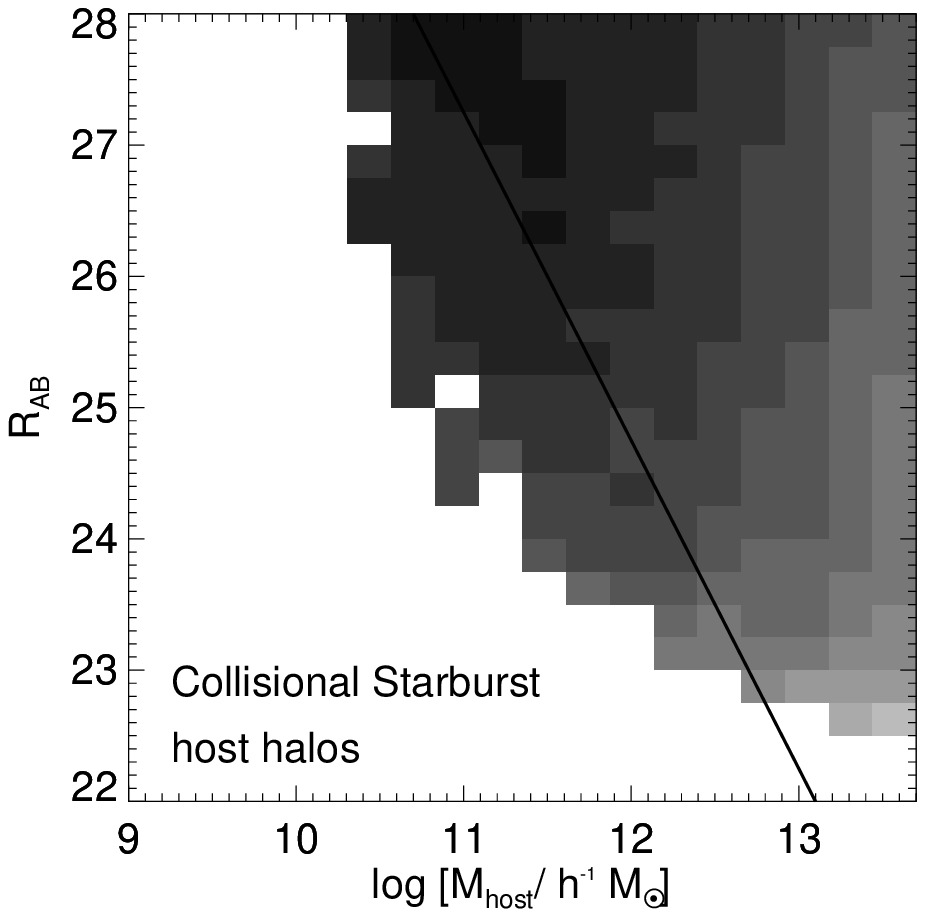}}
\caption{Joint probability distribution of extinction corrected
observed-frame ${\cal R}_{AB}$ magnitude and \emph{host} halo mass,
for the Constant Efficiency Quiescent (left) and Collisional Starburst
(right) models.  The shadings correspond to logarithmically spaced
density bins, and the line indicates a linear relation between mass
and luminosity.
\label{fig:hostmags}}
\end{figure*}

The relevant quantity for determining the correlation length, however,
is the mass of the virialized host halo containing the galaxies. The
joint distribution of galaxy magnitude and host halo mass is shown in
Figure~\ref{fig:hostmags}. This figure shows that in both models,
massive halos can host a number of galaxies of varying
luminosities. There is a critical luminosity, which reflects the
brightest galaxy that can be produced in a halo of a given mass, and
which is a fairly strong function of halo mass. The resulting weak
dependence of luminosity on host halo mass, however, is not sufficient
to produce a strong trend in the clustering strength with luminosity.
The weak dependence of clustering on luminosity, which arises from a
similar effect, has been noted before for galaxies at $z=0$ in
semi-analytic models \citep{slsdkw:00}.

Thus we argue that the weak dependence of clustering on luminosity is
a generic feature of these types of hierarchical models, {\em whether
or not they include a bursting mode of star formation}. Therefore,
this test does not provide as strong a constraint on star formation
modeling as we might have hoped, but rather is a reflection of the
fact that significant sub-structure is present in halos.  

We point out, however, that the scatter in the luminosity of objects
versus the host mass is sensitive to the subhalo multiplicity function
as determined by our semi-analytic models.  If the number of low-mass
subhalos per host were reduced, then the scatter in luminosity at
fixed host mass would also be reduced, producing a stronger dependence
of clustering on luminosity.  Indeed, when the satellite multiplicity
function from the semi-analytic models is compared with the subhalo
multiplicity function obtained from the ART simulations discussed in
\S \ref{sec:sims:ch}, we find that, for a fixed circular velocity, the
semi-analytic models produce a much larger number of subhalos per
massive halo.  This result may reflect the fact that the process of 
tidal disruption has been neglected in our semi-analytic treatment (see
\citealt{bkw:00}).  However, it is possible that the ART simulations
could overestimate the severity of subhalo destruction, which might be
reduced by the presence of condensed baryons (\citealt{khw:99} find that
the correlation length of $z=3$ galaxies identified in their
hydrodynamic simulations depends only very weakly on baryonic mass or
number density, in agreement with our results).  We defer a more
detailed investigation of this issue to a later work \citep{wech:01}.

\section{RELATING HALO COLLISIONS TO STARBURST GALAXIES}
\label{sec:compare}

In \S \ref{sec:sims:occ}, we showed the halo occupation number as a
function of host halo mass for the Colliding Halo model and for the
semi-analytic Collisional Starburst model.  Although these two models
represent the same physical scenario, i.e., one in which most bright
galaxies at high redshift are the product of a collision-triggered
burst of star formation, the results for the number of objects as a
function of mass in the two models were quite different.  If the
number of objects is modeled by a power-law function of the mass of
the host halo, we find that the slope of the occupation function for
collisions identified in the simulation ($S\simeq 1.1$) is steeper
than that of the observable galaxies produced in the semi-analytic
Collisional Starburst model ($S=0.7$). In this section, we attempt to
understand the source of this difference, and examine in detail the
importance of various aspects of the recipe used to model starbursts
in the semi-analytic model.  This section is rather detailed, and may
be skipped by the casual reader.

There are two possible causes for the discrepancy. Either the merger
rate in the semi-analytic models disagrees with the merger rate
measured from the simulations, or the difference is produced by the
more detailed semi-analytic treatment of the luminosity of the burst
resulting from each merger. 

Clearly, one expects the simulations to do the most accurate job of
properly identifying halo (and subhalo) collisions, at least above
their resolution limit, because this is dependent solely on how matter
interacts via gravity, which the simulation clearly represents more
accurately than a semi-analytic model.  However, it is possible that
mergers below the resolution limit of the simulation could produce
observable galaxies. Only halos with modeled mass of at least 50
particles ($6.25\times 10^{9} \hMsun$) are included in the halo
catalog, and it is estimated to be 100\% complete for masses above
about $2\times 10^{10} \msun$ \citep{sigad:00}.

The semi-analytic model can be run with arbitrarily high resolution;
in practice the trees are truncated at halos with circular velocities of
40 km/s, which corresponds to a mass resolution of roughly $1\times
10^8 \hMsun$ at $z=3$. The merger rate of galaxies (subhalos) is
modeled using several approximations: extended Press--Schechter is used
to construct the merger trees \citep{sk:99}, and the merging of
subhalos within virialized halos is modeled via the dynamical friction
and modified mean free path approximations (see
\S\ref{sec:sims:sams}).  Each of these approximations have been tested
in isolation (see \citealt{kolatt:00} for a recent analysis), but it
is unknown how accurately the merger rate produced by the whole
machinery agrees with simulations.  An additional concern is that the
\emph{definition} of what constitutes a merger may differ between the
semi-analytic models and the simulations.

\begin{figure*}
\resizebox{0.47\textwidth}{!}{\includegraphics{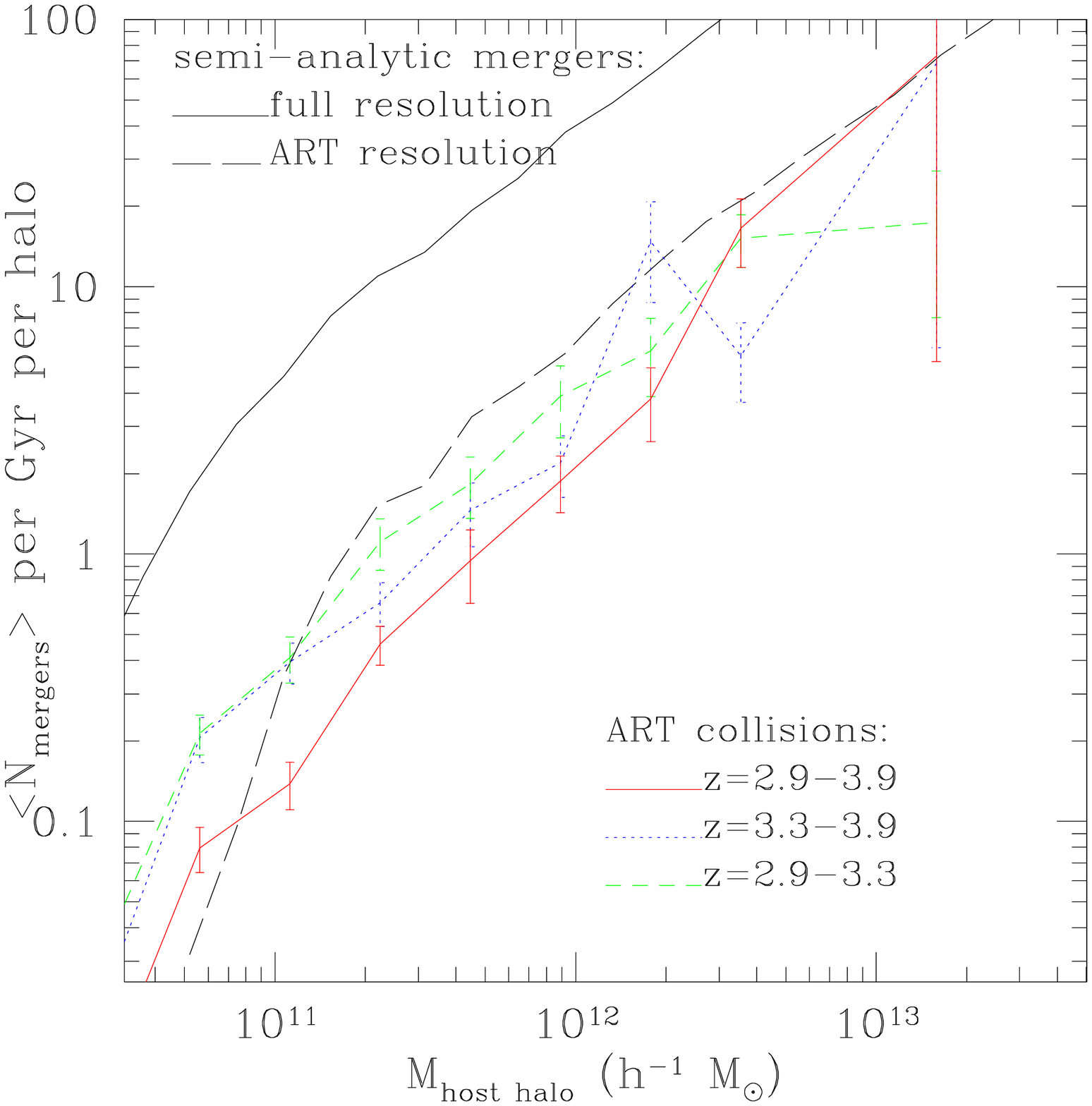}}
\resizebox{0.47\textwidth}{!}{\includegraphics{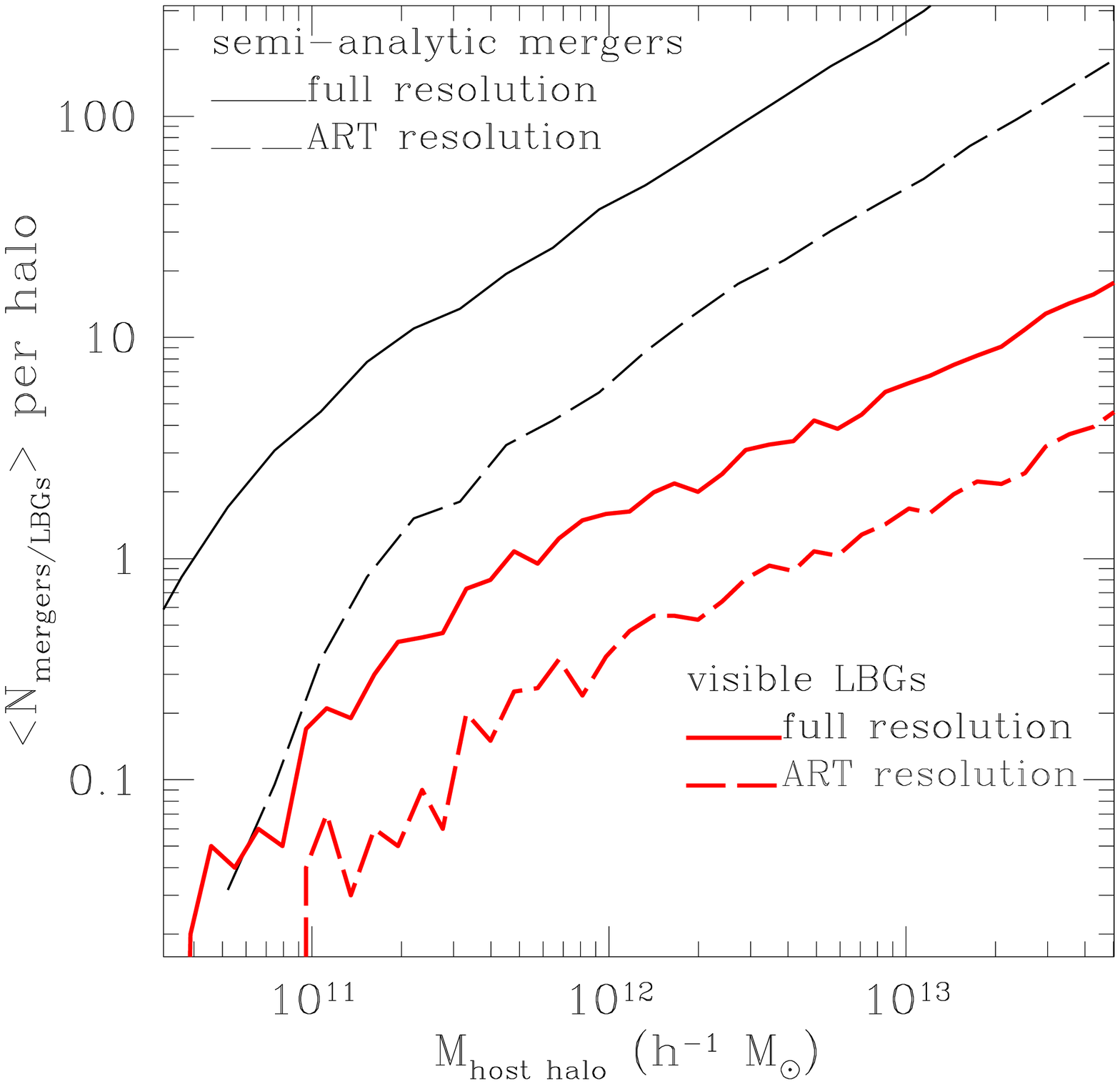}}
\figcaption{(Left) Average number of mergers per halo over the
redshift interval $z=2.9-3.9$, as a function of host halo mass (see
text for detailed definition), for the semi-analytic model and the ART
simulation (includes unbound collisions).  The dashed line shows only
those mergers in the semi-analytic model for which each merging halo
is above the resolution limit of the simulation; the solid line shows
all the semi-analytic mergers.  (Right) Average number of mergers per
halo over the redshift interval $z=2.9-3.9$ in the semi-analytic
model, and the average number of LBGs at $z=2.9$ in the CSB model
(bold). Solid lines show all mergers at the full resolution of the
semi-analytic model; dashed lines show only those mergers that would
be identified with the resolution of the ART simulation.
\label{fig:nm}}
\end{figure*}

In Figure \ref{fig:nm} (left panel), the number of mergers per host
halo as a function of the host mass measured in the ART simulations is
compared with the same quantity estimated in the semi-analytic model.
Both the {\em total} number of semi-analytic mergers and the
semi-analytic mergers assuming the completeness function of the
simulations (\citealt{sigad:00} are shown; the latter is equivalent to
imposing the mass resolution of the simulation onto the semi-analytic
models).  In the simulations, all collisions that occur during some
high-redshift timestep interval are identified, and assigned to the
distinct (i.e., non-sub) ``host'' halos that they reside in at a later
redshift.  A similar thing is done for the semi-analytic model to make
a comparison: we identify all mergers in the model that occur within
the same timestep interval, and assign them to the host halo that they
end up in at the end of the timestep.  Although the actual number of
mergers changes considerably as a function of assumed resolution in
the semi-analytic models, the \emph{shape} of the occupation function
doesn't change with resolution.  We have also tested the effects of
resolution directly by comparing the results of this simulation with
the analysis of a larger box with 1/8 the mass resolution, and find a
similar result.  The semi-analytic results match the simulation within
the (rather large) errors, although the slope is slightly shallower
than the best-fit power-law from the simulation.  The normalization is
not entirely consistent, however, there are many possible reasons for
this discrepancy --- as has been discussed --- and since the
normalization is fixed for these models by comparison with
observations and we are mainly concerned with the slope, this will not
affect the results.

Inaccuracies in the merger rate built into the semi-analytic models
therefore do not seem to be responsible for the discrepancy. We now
examine the ingredients of the recipe for assigning luminosities to
the mergers and determine how this affects the results.  In Figure
\ref{fig:nm} (right panel), the two lines from the left panel of
the figure are repeated, showing the number of mergers in the
semi-analytic model over the redshift interval $2.9 < z < 3.9$, for
the full resolution and with the ART resolution imposed. For
comparison, we show on the same panel the number of LBGs that would be
``observable'' (as usual, defined here as galaxies with ${\cal R}_{AB}
\leq 25.5$) in the semi-analytic model, both for the full resolution
and for the case in which the model has the same resolution as the
simulations.  Two things are apparent: first, there are a large number
of galaxies that would be bright enough to be included in our
``Steidel-like'' sample, and that are produced by mergers below the
mass resolution of the ART simulation\footnote[5]{In K99 it was argued
that the mass resolution of the ART simulation was adequate to model
all objects that would be observable in a Steidel-like sample. The
discrepancy between that argument and the semi-analytic results is
mainly due to the assumed dependence of burst efficiency on the mass
ratio of the mergers (including what assumption is made about the
minimum mass ratio that can produce a visible galaxy), and to
differences in the assignment of gas masses to halos.}, and second,
the resolution does not affect the slope of the occupation function
for galaxies.  The mergers in the semi-analytic model show a
significantly steeper increase with host halo mass than the {\em
observable galaxies} in the same model (virtually all of which were
made bright by recent mergers), indicating that for some reason a
collision is less likely to produce a bright galaxy if it occurs in a
massive halo.

We now investigate which elements of the semi-analytic recipes produce
this effect (Figure~\ref{fig:lumcomp}).  First we consider a simple
recipe for assigning luminosities to halo mergers, similar to that
used in \citet{kolatt:99}.  We assume that before each collision every
galaxy has a cold gas reservoir that is a constant fraction of the
(galactic) halo mass ($m_{\rm g} = f_{\rm g} f_{\rm b} m_{\rm halo}$,
where $f_{\rm b} \equiv \Omega_{\rm b} \Omega^{-1}_{\rm m}$ is the
fraction of mass in baryons and $f_{\rm g}$ is the fraction of baryons
in cold gas).  The mergers are divided into major ($m_2/m_1 > 0.25$)
and minor mergers, and every collision is assumed to produces a burst
of duration $\tau_{\rm burst}= 50$ Myr, during which 75\% and $50\%$
of the gas is converted into stars for major and minor mergers
respectively.  We assume that the mergers are uniformly distributed
over the timestep.  The apparent rest-1600 \AA\ magnitude of each
burst is estimated at the end of the timestep ($z=2.9$), using
Bruzual-Charlot (GISSEL00) stellar-population synthesis models
(assuming solar metallicity and a Salpeter initial mass function).
This recipe ($f_{gas}$ = C, Kolatt
\etal\ efficiency) is applied to the recorded mergers from the
semi-analytic model. Comparing the resulting number of observable
galaxies with the total number of mergers in Figure~\ref{fig:lumcomp},
we see that not all of the mergers produce observable galaxies, but
the galaxy occupation function is actually even \emph{steeper} than
the mergers. This is not surprising, as we have assumed that a
constant fraction of the halo mass is in the form of cold gas, so
massive halos have more gas and are more likely to produce bright
objects.

There are, however, a number of differences between this simple
prescription and the full treatment of the semi-analytic model. The
most relevant aspects and their treatment in the semi-analytic model
are summarized below:

\begin{itemize}
\item {\it Cold gas supply:} depends on halo mass and collapse time,
whether the galaxy is a central or satellite galaxy, and consumption by
previous star formation and expulsion by supernovae feedback.
\item {\it Burst efficiency:} modeled as a function of the mass ratio
and morphology of the colliding galaxies.  The efficiency of bursts in
major mergers is nearly independent of morphology, but bursts in minor
mergers are suppressed when a bulge is present.
\item {\it Burst timescale:} modeled as equal to the dynamical time of
the disk. 
\end{itemize}
We discuss each of these in turn.  

One can imagine that the more detailed modeling of the cold gas supply
might go in the right direction. More massive halos have a much lower
fraction of their mass in the form of cold gas, because the time for
the gas to cool out to the virial radius is larger than a Hubble time.
In addition, large halos will have many satellite galaxies, which are
not allowed to receive any new gas from cooling, and therefore exhaust
or expel their gas supply through star formation and supernovae
winds. Both effects might lead to fainter bursts in massive halos
because of a shortage of cold gas. To test this, for each merger in
the semi-analytic model, we record the gas content of both
progenitors. We now use this (SPF $f_{gas}$) instead of the constant gas
fraction assumed above, but leave the other ingredients the same, and
compute the number of observable galaxies as before. As is shown in 
Figure~\ref{fig:lumcomp}, some of the bursts in massive halos are
suppressed by the gas supply effect, but the slope of the occupation
function remains steeper than the full model in the largest mass host
halos.

\begin{inlinefigure}
\begin{center}
\resizebox{\textwidth}{!}{\includegraphics{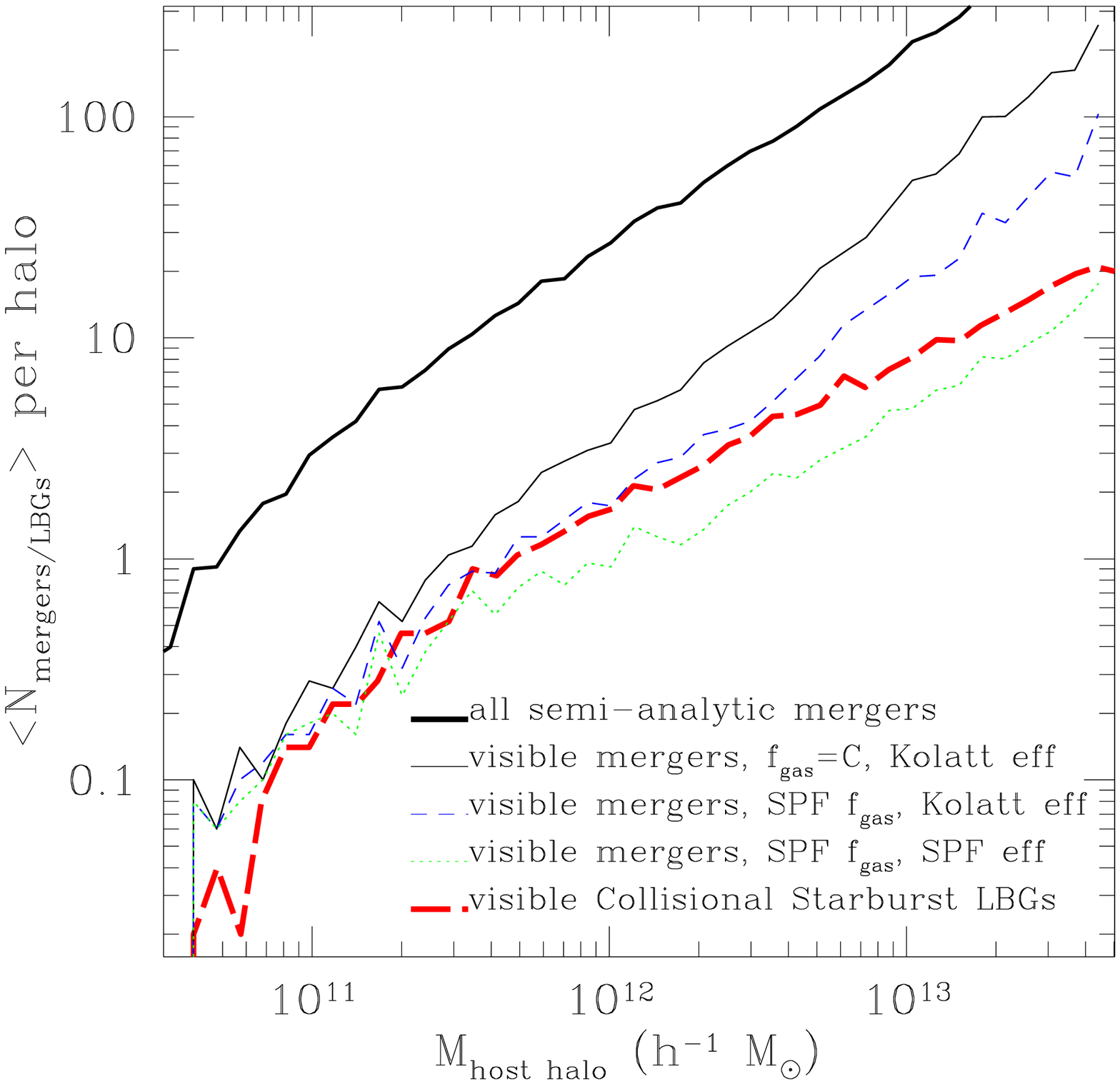}}
\end{center}
\figcaption{
Average number of mergers per halo, from $z=2.9-3.9$ for the semi-analytic
model, compared with galaxies with ${\cal R}_{AB} \leq 25.5$ (no dust correction)
in the same model at $z=2.9$, where luminosities have been assigned using:
a) the simple recipe of K99,
b) same as a) but using the gas contents of the full CSB model, 
c) same as b) but using the bulge-fraction dependent burst 
efficiency function of SPF, and
d) the actual CSB model.
\label{fig:lumcomp}}
\end{inlinefigure}

Next we try the burst efficiency recipe of SPF (see
Eqn.~\ref{eqn:eburst}), including the dependence on the bulge
fraction.  Using this prescription, the occupation function for
observable galaxies agrees fairly well with the results of the full
model --- at least the slope at the high-mass end is the same. The
total number of galaxies is a bit smaller than in the full model, but
this is perhaps to be expected as we have neglected quiescent star
formation in this simple exercise. Note that in these models, bulges
are built up by major mergers. Therefore it is not surprising that the
massive halos, which formed from higher peaks in the initial density
field, are more likely to contain galaxies with prominent bulges ---
this is just the high-redshift analog of the morphology-density
relationship. The suppression of bursts in minor mergers for galaxies
with existing bulges appears to be the main effect that flattens the
LBG occupation function in the full model.

The effects investigated above seem to account for the differences
between the simple Colliding Halo model and the full semi-analytic
Collisional Starburst model, but this exercise has highlighted the
sensitivity of our results to the detailed modeling of the efficiency
and time dependence of star formation in the collisional starbursts,
which remain highly uncertain. In particular, because minor mergers
are so much more common than major mergers, the treatment of bursts in
minor mergers is very important --- yet it is also more sensitive to the
details of the interaction.  We are working to better understand these
issues using a set of hydrodynamic simulations of colliding galaxies,
similar to those of \citet{mihos:94}, but with initial conditions
chosen to be representative of $z \sim 3$ galaxies
\citep{somerville:00}, and covering a larger parameter space.

\section{DISCUSSION AND CONCLUSIONS}
\label{sec:conclu}

We have investigated a range of models, spanning the most extreme
versions of previously proposed scenarios for Lyman-break galaxies,
and compared the predicted clustering properties of galaxies at $z=3$
with those of Lyman-break galaxies with spectroscopic redshifts from
the ground-based sample of Steidel, Adelberger, and collaborators. We
investigated two simple models for assigning observable galaxies to
halos, the Massive Halo model, which associates one LBGs with each
massive halo, and the Colliding Halo model, which associates an
LBG with each halo collision. In addition, we investigated three
``realistic'' models based on the semi-analytic models of SPF, which are
differentiated by their assumptions about the efficiency of
transforming cold gas into stars.

All five models are normalized to produce the observed, incompleteness
corrected number density of LBGs \citep{steidel:99}.  For each model,
we then compute the halo occupation function, or the number of
observable galaxies per halo as a function of the host halo mass.
These occupation functions are then used to populate dark-matter halos
in a large, dissipationless \LCDM\ N-body simulation with galaxies. We
select galaxies, attempting to mimic the observational selection
techniques, and then use the resulting catalogs to calculate the
clustering properties of LBGs in each scenario.

Clustering statistics that smooth over a relatively large region (such
as the overdensity distribution in cells of $\sim 12$ \hmpc\ on a
side, or the correlation length) are fairly non-discriminatory for all
five models.  No models can be strongly ruled out based on the
correlation length; the most clustered is the Colliding Halo model,
which is about $2\sigma$ away from the data.  The correlation lengths
for the rest of the models are in reasonable agreement with the data,
when compared using the same technique.  However, there is still
uncertainty in the determination of the true correlation length of
LBGs; if it is really closer to $3$ \hmpc\, rather than 4--5 \hmpc,
then this is probably an indication that LBGs must be hosted by
smaller mass halos than any of our models predict.  A change of a
factor of two in the true number density of the objects seen as LBGs
could possibly account for this. Alternatively, for the ``realistic''
models, the uncertainties in the modeling of star formation, feedback,
and dust extinction, the stellar initial mass function, etc., are
likely to be sufficient to allow for some adjustments that would
change the occupation function in just the right way to satisfy all
the constraints investigated here.  One possibility, which we have not
investigated in detail here, is that the most rapidly star-forming
galaxies suffer so much dust extinction that they are not included in
the observational sample.  In this case, the proportion of visible
galaxies in massive halos would be further reduced --- thus lowering
the correlation length.  The hope, of course, is that the empirical
understanding gained by this sort of schematic modeling will lead to
constraints that will eventually bring about progress towards a deeper
understanding of these physical processes.

The fraction of galaxies in close pairs, which is a good measure of
the number of multiple galaxies in a single halo, proved to be a
better discriminator between models than the statistics that measure
clustering on larger scales. This is because this quantity is quite
sensitive to the slope $S$ of the halo occupation function at the high
mass end, $N_g \propto M^S$.  This statistic can discriminate between
the most extreme models, the simple Massive Halo model ($S=0$), which
underproduces close pairs by more than $1\sigma$, and the Colliding
Halo model ($S=1.1$), which overproduces them by several $\sigma$ (see
Figure \ref{fig:pairs}).  The Collisional Starburst model, which has a
halo occupation slope of $S\sim 0.7$ produces good agreement with the
close pair fraction of the observations; the other two ``realistic''
semi-analytic models yield a slightly higher slope of $S\sim 0.8$, and
a higher fraction of pairs than the data, but are not strongly ruled
out by the current data.

We also examined the dependence of clustering on luminosity. There
have been suggestions that the correlation length derived from the
observations depends on the magnitude limit of the sample, with
brighter galaxies being more strongly clustered (S98, G00). This has
been interpreted as evidence for a tight connection between halo mass
and galaxy luminosity, which would seem to disfavor models with
stochastic star formation such as collisional starbursts. However,
when all available measurements of the correlation length from the
literature are compiled, it seems clear that the strength of this
observational trend depends greatly on which subset of the
observations one chooses to include.  Until a single high-redshift
sample exists that is large enough in volume to measure the clustering
of bright galaxies without being dominated by shot-noise, and deep
enough to simultaneously measure the clustering of much fainter
galaxies, it will be difficult to place strong limits on the models.

In the meantime, for the models, we find that \emph{both} the
quiescent and collisional starburst models display only a weak trend
of correlation length with luminosity, much weaker than that suggested
by the observational analysis of G00, or predicted by the Massive Halo
model. As expected, the relation between galaxy luminosity and halo
mass has much larger scatter in starburst models than in quiescent
models.  However, since massive halos can have a number of subhalos,
with a variety of masses, both models actually have quite a large
scatter in the luminosities of galaxies that reside in these host
halos, which are what determine the clustering properties.

Any hierarchical clustering model will have multiple subhalos within
the massive halos, with a range of masses, several of which are likely
to correspond to observable galaxies. In fact, the observed fraction
of galaxies in close pairs seems to indicate that many of the halos do
host more than one visible galaxy, since the number of pairs in the
Massive Halo model is more than $1\sigma$ lower than the data (the
first bin in Figure~\ref{fig:pairs}). It is a concern that there is
some disagreement between the subhalo multiplicity function of the
semi-analytic models compared to that in the simulations; if in fact
the number of subhalos predicted by the semi-analytic treatment is too
high, the dependence of clustering on luminosity would be somewhat
stronger.

We found that two versions of the collisional starburst scenario gave
different results for the halo occupation function and tried to
understand why. The Colliding Halo model, based on properties of
collisions identified in high-resolution N-body simulations, had many
more LBGs in massive halos than the semi-analytic model, and a steeper
halo occupation function.  Although the difference could have been
caused by an inaccurate treatment of halo mergers in the semi-analytic
analysis, we found that this was not the case.  Instead, the
discrepancy between this model and the semi-analytic Collisional Starburst
model arises because of the detailed modeling of gas processes
and starburst efficiency in the latter.  Namely,
high-mass halos in the semi-analytic model are less efficient at
producing bright galaxies.  This is primarily due to two effects:
massive halos collapse more recently and have not had time to cool as
large a fraction of their gas, and (the more dominant effect) bursts
in minor mergers are suppressed when the primary galaxy already
contains a prominent bulge. This recipe was adopted by SPF to attempt
to capture the behavior found by \citet{mihos:96}, based on
hydrodynamic simulations of merging galaxies. Apparently, a
morphology-density relation is already in place in our models at $z
\sim 3$, and this has a significant impact on the predicted properties
of observable galaxies. It should be noted that although SPF based
their recipe for burst efficiency on the best simulations that were
available at the time, the sensitivity of our results to this
ingredient is a concern. A more extensive set of simulations, with
initial conditions chosen to better represent $z \sim 3$ galaxies and
covering a larger region of parameter space, is in progress
\citep{somerville:00} and will hopefully improve our
understanding of this process.  In the meantime, one should be aware
that the predictions for clustering in this model are especially
sensitive to the assumed efficiency of converting gas into stars
during a starburst.

In summary, although one might have expected the clustering properties
of galaxies on intermediate scales (i.e., the correlation length) to
provide a strong discriminator between models of galaxy formation, in
fact we find that even extreme models yield similar results for this
statistic --- although models with a very steep occupation function
are only marginally acceptable.  Clustering statistics that probe
smaller scales are a better way to discriminate between models whose
halo occupation slopes are different, but even with this statistic,
none of the more realistic models can be strongly ruled out with the
data sample used here.  Interestingly, we found that the halo
occupation slope was shallower in the Collisional Starburst model than
in the massive quiescent models investigated, laying to rest concerns
that this model would comparatively have too many close pairs.  When
combined with the failings of the other two (quiescent) models
discussed in SPF, it still appears that the Collisional Starburst
provides the best agreement with all the available data.  Still, we
emphasize that many ingredients of the modeling of collisional
starbursts remain highly uncertain, and if indeed this process is
responsible for producing most of the galaxies observed at high
redshift, further investigation will be crucial.  It should also be
emphasized that although the current observational situation is still
too uncertain to unambiguously determine how these high-redshift
galaxies are related to the underlying dark halos, there is hope that
in the future, a combination of observations on number density, and
both small-scale and large-scale clustering, should be able to
determine the halo occupation function, which we then can hope to
explain with physical models for galaxy formation.

\acknowledgments
\section{ACKNOWLEDGMENTS}
We thank Kurt Adelberger for help with interpreting the data, for
providing some data in electronic form, and for several useful
discussions throughout the course of this project.  We also thank
Sandy Faber, Mauro Giavalisco, Lars Hernquist, Patrik Jonsson, and Ari
Maller for useful discussions and comments.  We would also like to
thank Andrey Kravtsov \& Anatoly Klypin for running and providing
access to the ART simulations, which were run at NSCA and NRL, and the
GIF collaboration for providing access to their simulations. RHW has
received support from a NSF GAANN fellowship at UCSC.  JRP
acknowledges a Humboldt Award at the Max Planck Institute for Physics
in Munich.  JSB received support from NASA LTSA grant NAG5-3525 and
NSF grant AST-9802568.  This work was also supported by a NASA ATP
grant, NSF grants AST-9529098 and PHY-9722146, and a Faculty Research
Grant at UCSC, and by the US-Israel Binational Science Foundation
grant 98-00217 and Israel Science Foundation grant 546/98.  We also
acknowledge the hospitality of the Institute for Theoretical Physics
at UC Santa Barbara, and RHW acknowledges the hospitality of the
Hebrew University of Jerusalem, the IoA, Cambridge, and the Max Planck
Institute for Astrophysics, Garching.

\bibliographystyle{apj}
\bibliography{risa}

\begin{thebibliography}{65}
\expandafter\ifx\csname natexlab\endcsname\relax\def\natexlab#1{#1}\fi

\bibitem[{{Adelberger}(2000)}]{adel:99}
{Adelberger}, K.~L. 2000, in ASP Conference Series, Vol. 200, Clustering at
  High Redshift, ed. A.~{Mazure}, O.~{Le Fevre}, \& V.~{Le Brun} (Kluwer
  Academic Publishers)

\bibitem[{{Adelberger} \& {Steidel}(2000)}]{as:00}
{Adelberger}, K.~L. \& {Steidel}, C.~C. 2000, \apj, 544, 218

\bibitem[{{Adelberger} {et~al.}(1998){Adelberger}, {Steidel}, {Giavalisco},
  {Dickinson}, {Pettini}, \& {Kellogg}}]{adel:98}
{Adelberger}, K.~L., {Steidel}, C.~C., {Giavalisco}, M., {Dickinson}, M.,
  {Pettini}, M., \& {Kellogg}, M. 1998, \apj, 505, 18, (A98)

\bibitem[{{Adelberger} {et~al.}(2001){Adelberger}, {Steidel}, {Giavalisco},
  {Dickinson}, {Pettini}, \& {Kellogg}}]{adel:01}
---. 2001, in preparation

\bibitem[{{Arnouts} {et~al.}(1999){Arnouts}, {Cristiani}, {Moscardini},
  {Matarrese}, {Lucchin}, {Fontana}, \& {Giallongo}}]{arnouts:99}
{Arnouts}, S., {Cristiani}, S., {Moscardini}, L., {Matarrese}, S., {Lucchin},
  F., {Fontana}, A., \& {Giallongo}, E. 1999, \mnras, 310, 540

\bibitem[{{Bagla}(1998)}]{bagla:98a}
{Bagla}, J.~S. 1998, \mnras, 297, 251

\bibitem[{{Baugh} {et~al.}(1999){Baugh}, {Benson}, {Cole}, {Frenk}, \&
  {Lacey}}]{baugh:99}
{Baugh}, C.~M., {Benson}, A.~J., {Cole}, S., {Frenk}, C.~S., \& {Lacey}, C.~G.
  1999, \mnras, 305, L21

\bibitem[{{Baugh} {et~al.}(1998){Baugh}, {Cole}, {Frenk}, \&
  {Lacey}}]{baugh:98}
{Baugh}, C.~M., {Cole}, S., {Frenk}, C.~S., \& {Lacey}, C.~G. 1998, \apj, 498,
  504

\bibitem[{{Benson} {et~al.}(2000){Benson}, {Cole}, {Frenk}, {Baugh}, \&
  {Lacey}}]{benson:00}
{Benson}, A.~J., {Cole}, S., {Frenk}, C.~S., {Baugh}, C.~M., \& {Lacey}, C.~G.
  2000, \mnras, 311, 793

\bibitem[{Blanton {et~al.}(2000)Blanton, Cen, Ostriker, Strauss, \&
  Tegmark}]{blanton:00}
Blanton, M., Cen, R., Ostriker, J., Strauss, M., \& Tegmark, M. 2000, \apj,
  531, 1

\bibitem[{{Bullock} {et~al.}(2000{\natexlab{a}}){Bullock}, {Kolatt}, {Sigad},
  {Somerville}, {Kravtsov}, {Klypin}, {Primack}, \& {Dekel}}]{bullock:00}
{Bullock}, J.~S., {Kolatt}, T.~S., {Sigad}, Y., {Somerville}, R.~S.,
  {Kravtsov}, A.~V., {Klypin}, A.~A., {Primack}, J.~R., \& {Dekel}, A.
  2000{\natexlab{a}}, \mnras, 321, 559

\bibitem[{{Bullock} {et~al.}(2000{\natexlab{b}}){Bullock}, {Kravtsov}, \&
  {Weinberg}}]{bkw:00}
{Bullock}, J.~S., {Kravtsov}, A.~V., \& {Weinberg}, D. 2000{\natexlab{b}},
  \apj, 539, 517

\bibitem[{Cole {et~al.}(1994)Cole, Arag\'{o}n-Salamanca, Frenk, Navarro, \&
  Zepf}]{cafnz:94}
Cole, S., Arag\'{o}n-Salamanca, A., Frenk, C., Navarro, J., \& Zepf, S. 1994,
  \mnras, 271, 781

\bibitem[{{Coles} {et~al.}(1998){Coles}, {Lucchin}, {Matarrese}, \&
  {Moscardini}}]{coles:98}
{Coles}, P., {Lucchin}, F., {Matarrese}, S., \& {Moscardini}, L. 1998, \mnras,
  300, 183

\bibitem[{Gardner {et~al.}(1999)Gardner, Katz, Hernquist, \&
  Weinberg}]{gardner:99}
Gardner, J., Katz, N., Hernquist, L., \& Weinberg, D. 1999, \apj, submitted
  (astro-ph/9911343)

\bibitem[{{Giavalisco} \& {Dickinson}(2001)}]{giav:00}
{Giavalisco}, M. \& {Dickinson}, M.~E. 2001, \apj, in press (astro-ph/0012249,
  G00)

\bibitem[{{Giavalisco} {et~al.}(1998){Giavalisco}, {Steidel}, {Adelberger},
  {Dickinson}, {Pettini}, \& {Kellogg}}]{giav:98}
{Giavalisco}, M., {Steidel}, C.~C., {Adelberger}, K.~L., {Dickinson}, M.~E.,
  {Pettini}, M., \& {Kellogg}, M. 1998, \apj, 503, 543, (G98)

\bibitem[{{Governato} {et~al.}(1998){Governato}, {Baugh}, {Frenk}, {Cole},
  {Lacey}, {Quinn}, \& {Stadel}}]{gov:98}
{Governato}, F., {Baugh}, C.~M., {Frenk}, C.~S., {Cole}, S., {Lacey}, C.~G.,
  {Quinn}, T., \& {Stadel}, J. 1998, \nat, 392, 359

\bibitem[{{Haiman} \& {Hui}(2001)}]{haiman:00}
{Haiman}, Z. \& {Hui}, L. 2001, \apj, in press (astro-ph/0002190)

\bibitem[{{Jenkins} {et~al.}(1998){Jenkins}, {Frenk}, {Pearce}, {Thomas},
  {Colberg}, {White}, {Couchman}, {Peacock}, {Efstathiou}, \&
  {Nelson}}]{jenk:98}
{Jenkins}, A., {Frenk}, C.~S., {Pearce}, F.~R., {Thomas}, P.~A., {Colberg},
  J.~M., {White}, S. D.~M., {Couchman}, H. M.~P., {Peacock}, J.~A.,
  {Efstathiou}, G., \& {Nelson}, A.~H. 1998, \apj, 499, 20

\bibitem[{{Jing}(1999)}]{jing:99}
{Jing}, Y.~P. 1999, \apjl, 515, L45

\bibitem[{{Jing} \& {Suto}(1998)}]{jing:98}
{Jing}, Y.~P. \& {Suto}, Y. 1998, \apjl, 494, L5

\bibitem[{{Katz} {et~al.}(1999){Katz}, {Hernquist}, \& {Weinberg}}]{khw:99}
{Katz}, N., {Hernquist}, L., \& {Weinberg}, D.~H. 1999, \apj, 523, 463

\bibitem[{{Kauffmann} {et~al.}(1999{\natexlab{a}}){Kauffmann}, {Colberg},
  {Diaferio}, \& {White}}]{kauf:99a}
{Kauffmann}, G., {Colberg}, J.~M., {Diaferio}, A., \& {White}, S. D.~M.
  1999{\natexlab{a}}, \mnras, 303, 188

\bibitem[{{Kauffmann} {et~al.}(1999{\natexlab{b}}){Kauffmann}, {Colberg},
  {Diaferio}, \& {White}}]{kauf:99b}
---. 1999{\natexlab{b}}, \mnras, 307, 529

\bibitem[{Kauffmann {et~al.}(1997)Kauffmann, Nusser, \& Steinmetz}]{kns:97}
Kauffmann, G., Nusser, A., \& Steinmetz, M. 1997, \mnras, 286, 795

\bibitem[{{Kennicutt}(1998)}]{kennicutt}
{Kennicutt}, R.~C. 1998, \apj, 498, 541

\bibitem[{Kolatt {et~al.}(2001)Kolatt, Bullock, Sigad, Primack, Dekel,
  Kravtsov, \& Klypin}]{kolatt:00}
Kolatt, T., Bullock, J., Sigad, Y., Primack, J., Dekel, A., Kravtsov, A., \&
  Klypin, A. 2001, \mnras, submitted (astro-ph/0010222)

\bibitem[{{Kolatt} {et~al.}(1999){Kolatt}, {Bullock}, {Somerville}, {Sigad},
  {Jonsson}, {Kravtsov}, {Klypin}, {Primack}, {Faber}, \& {Dekel}}]{kolatt:99}
{Kolatt}, T.~S., {Bullock}, J.~S., {Somerville}, R.~S., {Sigad}, Y., {Jonsson},
  P., {Kravtsov}, A.~V., {Klypin}, A.~A., {Primack}, J.~R., {Faber}, S.~M., \&
  {Dekel}, A. 1999, \apjl, 523, L109, (K99)

\bibitem[{{Kravtsov} {et~al.}(1997){Kravtsov}, {Klypin}, \&
  {Khokhlov}}]{kkk:97}
{Kravtsov}, A.~V., {Klypin}, A.~A., \& {Khokhlov}, A.~M. 1997, \apjs, 111, 73

\bibitem[{{Kuiper}(1962)}]{kuiper:62}
{Kuiper}, N.~H. 1962, in Proceedings of the Korinklijke Nederlandse Akademie
  van Wetenschappen, ser. A, Vol.~63, 38--47

\bibitem[{{Lacey} \& {Cole}(1993)}]{lacey:93}
{Lacey}, C. \& {Cole}, S. 1993, \mnras, 262, 627

\bibitem[{{Lowenthal} {et~al.}(1997){Lowenthal}, {Koo}, {Guzman}, {Gallego},
  {Phillips}, {Faber}, {Vogt}, {Illingworth}, \& {Gronwall}}]{low:97}
{Lowenthal}, J.~D., {Koo}, D.~C., {Guzman}, R., {Gallego}, J., {Phillips},
  A.~C., {Faber}, S.~M., {Vogt}, N.~P., {Illingworth}, G.~D., \& {Gronwall}, C.
  1997, \apj, 481, 673

\bibitem[{{Madau} {et~al.}(1996){Madau}, {Ferguson}, {Dickinson}, {Giavalisco},
  {Steidel}, \& {Fruchter}}]{madau:96}
{Madau}, P., {Ferguson}, H., {Dickinson}, M., {Giavalisco}, M., {Steidel}, C.,
  \& {Fruchter}, A. 1996, \mnras, 283, 1388

\bibitem[{{Magliocchetti} \& {Maddox}(1999)}]{mag:99}
{Magliocchetti}, M. \& {Maddox}, S.~J. 1999, \mnras, 306, 988

\bibitem[{{Makino} \& {Hut}(1997)}]{mh:97}
{Makino}, J. \& {Hut}, P. 1997, \apj, 481, 83

\bibitem[{Martini \& Weinberg(2001)}]{martini:00}
Martini, P. \& Weinberg, D. 2001, \apj, in press (astro-ph/0002384)

\bibitem[{Mihos \& Hernquist(1994)}]{mihos:94}
Mihos, J. \& Hernquist, L. 1994, \apjl, 425, L13

\bibitem[{Mihos \& Hernquist(1995)}]{mihos:95}
---. 1995, \apj, 448, 41

\bibitem[{Mihos \& Hernquist(1996)}]{mihos:96}
---. 1996, \apj, 464, 641

\bibitem[{{Mo} \& {Fukugita}(1996)}]{mofuku:96}
{Mo}, H.~J. \& {Fukugita}, M. 1996, \apjl, 467, L9

\bibitem[{{Mo} {et~al.}(1999){Mo}, {Mao}, \& {White}}]{mmw:99}
{Mo}, H.~J., {Mao}, S., \& {White}, S. D.~M. 1999, \mnras, 304, 175

\bibitem[{{Mo} \& {White}(1996)}]{mw:96}
{Mo}, H.~J. \& {White}, S. D.~M. 1996, \mnras, 282, 347

\bibitem[{{Moscardini} {et~al.}(1998){Moscardini}, {Coles}, {Lucchin}, \&
  {Matarrese}}]{moscar:98}
{Moscardini}, L., {Coles}, P., {Lucchin}, F., \& {Matarrese}, S. 1998, \mnras,
  299, 95

\bibitem[{{Peebles}(1980)}]{peebles:80}
{Peebles}, P. J.~E. 1980, The Large-Scale Structure of the Universe (Princeton
  University Press)

\bibitem[{{Press} \& {Schechter}(1974)}]{ps:74}
{Press}, W.~H. \& {Schechter}, P. 1974, \apj, 187, 425

\bibitem[{{Press} {et~al.}(1992){Press}, {Teukolsky}, {Vetterling}, \&
  {Flannery}}]{press:92}
{Press}, W.~H., {Teukolsky}, S.~A., {Vetterling}, W.~T., \& {Flannery}, B.~P.
  1992, Numerical Recipes in C (Cambridge University Press)

\bibitem[{{Sheth} \& {Tormen}(1999)}]{st:99}
{Sheth}, R.~K. \& {Tormen}, G. 1999, \mnras, 308, 119

\bibitem[{{Sigad} {et~al.}(2001){Sigad}, {Kolatt}, {Bullock}, {Kravtsov},
  {Klypin}, {Primack}, \& {Dekel}}]{sigad:00}
{Sigad}, Y., {Kolatt}, T., {Bullock}, J., {Kravtsov}, A.~V., {Klypin}, A.~A.,
  {Primack}, J.~R., \& {Dekel}, A. 2001, \mnras, submitted (astro-ph/0005323)

\bibitem[{Somerville(1997)}]{rsthesis}
Somerville, R. 1997, PhD thesis, University of California, Santa Cruz

\bibitem[{{Somerville} \& {Kolatt}(1999)}]{sk:99}
{Somerville}, R.~S. \& {Kolatt}, T.~S. 1999, \mnras, 305, 1

\bibitem[{{Somerville} {et~al.}(2000{\natexlab{a}}){Somerville}, {Lemson},
  {Kolatt}, \& {Dekel}}]{somerville:00}
{Somerville}, R.~S., {Lemson}, G., {Kolatt}, T.~S., \& {Dekel}, A.
  2000{\natexlab{a}}, \mnras, 316, 479

\bibitem[{Somerville {et~al.}(2001)Somerville, Lemson, Sigad, Dekel, Kauffmann,
  \& White}]{slsdkw:00}
Somerville, R.~S., Lemson, G., Sigad, Y., Dekel, A., Kauffmann, G., \& White,
  S. 2001, \mnras, 320, 289

\bibitem[{{Somerville} {et~al.}(2000{\natexlab{b}}){Somerville}, {Primack}, \&
  {Faber}}]{spf:00}
{Somerville}, R.~S., {Primack}, J.~R., \& {Faber}, S.~M. 2000{\natexlab{b}},
  \mnras, in press, (SPF)

\bibitem[{Spinrad {et~al.}(1998)Spinrad, Stern, Bunker, Dey, Lanzetta, Yahil,
  Pascarelle, \& Fernandez-Soto}]{spinrad:98}
Spinrad, H., Stern, D., Bunker, A., Dey, A., Lanzetta, K., Yahil, A.,
  Pascarelle, S., \& Fernandez-Soto, A. 1998, \aj, 116, 2617

\bibitem[{{Steidel} {et~al.}(1998){Steidel}, {Adelberger}, {Giavalisco},
  {Dickinson}, {Pettini}, \& {Kellogg}}]{steidel:98rs}
{Steidel}, C., {Adelberger}, K., {Giavalisco}, M., {Dickinson}, M., {Pettini},
  M., \& {Kellogg}, M. 1998, in Royal Society Discussion Meeting, "Large Scale
  Structure in the Universe", (astro-ph/9805267, S98)

\bibitem[{Steidel {et~al.}(1996{\natexlab{a}})Steidel, Giavalisco, Dickinson,
  \& Adelberger}]{steidel:96a}
Steidel, C., Giavalisco, M., Dickinson, M., \& Adelberger, K.
  1996{\natexlab{a}}, \aj, 112, 352

\bibitem[{Steidel {et~al.}(1996{\natexlab{b}})Steidel, Giavalisco, Pettini,
  Dickinson, \& Adelberger}]{steidel:96b}
Steidel, C., Giavalisco, M., Pettini, M., Dickinson, M., \& Adelberger, K.
  1996{\natexlab{b}}, \apjl, 462, L17

\bibitem[{Steidel \& Hamilton(1992)}]{steidel:92}
Steidel, C. \& Hamilton, D. 1992, \aj, 104, 941

\bibitem[{{Steidel} {et~al.}(1999){Steidel}, {Adelberger}, {Giavalisco},
  {Dickinson}, \& {Pettini}}]{steidel:99}
{Steidel}, C.~C., {Adelberger}, K.~L., {Giavalisco}, M., {Dickinson}, M., \&
  {Pettini}, M. 1999, \apj, 519, 1

\bibitem[{{Wang} \& {Heckman}(1996)}]{wh:96}
{Wang}, B. \& {Heckman}, T.~M. 1996, \apj, 457, 645

\bibitem[{{Wechsler}(2001)}]{wech:01}
{Wechsler}, R.~H. 2001, PhD thesis, University of California, Santa Cruz

\bibitem[{{Wechsler} {et~al.}(1998){Wechsler}, {Gross}, {Primack},
  {Blumenthal}, \& {Dekel}}]{wech:98}
{Wechsler}, R.~H., {Gross}, M. A.~K., {Primack}, J.~R., {Blumenthal}, G.~R., \&
  {Dekel}, A. 1998, \apj, 506, 19

\bibitem[{Weinberg {et~al.}(2000)Weinberg, Hernquist, \& Katz}]{weinberg:00}
Weinberg, D., Hernquist, L., \& Katz, N. 2000, \apj, submitted
  (astro-ph/0005340)

\bibitem[{Weymann {et~al.}(1998)Weymann, Stern, Bunker, Spinrad, Chaffee,
  Thompson, \& Storrie-Lombardi}]{weymann:98}
Weymann, R., Stern, D., Bunker, A., Spinrad, H., Chaffee, F., Thompson, R., \&
  Storrie-Lombardi, L. 1998, \apj, 505, L95

\end{thebibliography}

\end{document}